\documentstyle{ioplppt}                
\begin{document}
\jl{1}
\title{From the Birkhoff-Gustavson normalization
to the Bertrand-Darboux integrability condition
}
\author{
Yoshio Uwano
}
\address{
Department of Applied Mathematics and Physics, Kyoto University
\\
Kyoto 606-8501, Japan
}
\begin{abstract}
The Bertrand-Darboux integrability condition for a certain class of
perturbed harmonic oscillators is studied from the viewpoint of the
Birkhoff-Gustavson(BG)-normalization: In solving an inverse problem
of the BG-normalization on computer algebra, it is shown that
if the perturbed harmonic oscillators with a homogeneous
{\it cubic}-polynomial potential and with a homogeneous
{\it quartic}-polynomial potential share the same BG-normal form up
to degree-$4$, then both oscillators satisfy the Bertrand-Darboux
integrability condition.
\end{abstract}
%
%
%
\ams{70K45, 70H06, 37J40, 37J40, 37K10}
\maketitle
\newcommand{\ps}{{\bf R}^2 \times {\bf R}^2}
\newcommand{\ovl}{\overline}
\newcommand{\supim}{\mbox{{\tiny image}}}
\newcommand{\image}{\mbox{image}}
\newtheorem{definition}{Definition}[section]
\newtheorem{theorem}[definition]{Theorem}
\newtheorem{lemma}[definition]{Lemma}
\newtheorem{conjecture}[definition]{Conjecture}
\newtheorem{remark}{Remark}
\section{Introduction}
The Bertrand-Darboux theorem is well known to provide
a necessary and sufficient condition for two-degree-of-freedom
natural Hamiltonian systems associated with the Euclidean metric
to admit an integral of motion
quadratic in momenta (Darboux 1901). Moreover, that condition
is necessary and sufficient for the natural Hamiltonian systems
to be separable in either Cartesian, polar, parabolic or elliptic
coordinates  (Marshall and Wojciechowski 1988). This theorem thereby
provide a sufficient condition for the complete integrability of
the natural Hamiltonian systems, which will be referred to as
the Bertrand-Darboux integrability condition (BDIC) in the present
paper. The BDIC has been studied repeatedly from various viewpoints;
the separation of variables (Marshall and Wojciechowski 1988,
Grosche et al 1995 in the path-integral formulation),
the complete integrability (Perelomov 1990),
and so-called the direct method (Hietarinta 1987), for example
(see also Whittaker 1944 as an older reference and
the references in the above-cited literatures).
\par
The aim of the present paper is to show that the BDIC is obtained
as an outcome of a study on the Birkhoff-Gustavson(BG)-normalization
(Gustavson 1966, Moser 1968) for the Hamiltonians
of the perturbed harmonic oscillators with homogeneous
{\it cubic}-polynomial potentials (PHOCP's).
In solving an {\it inverse problem} of the BG-normalization for
a given PHOCP-Hamiltonian with a help of computer algebra,
the family of the perturbed harmonic oscillators
with homogeneous {\it quartic}-polynomial potentials (PHOQP's)
is identified which share the same BG-normal form
up to degree-$4$ with the given PHOCP.
Consequently, a new deep relation is found between
the BDIC for the PHOCP's and that for the PHOQP's:
It is shown that if a PHOCP and a PHOQP
share the same BG-normal form up to degree-$4$ then both oscillators
are integrable in the sense that they satisfy the BDIC.
It is worth noting that the present work 
was inspired in the debugging process of the computer-programme named
\lq ANFER' for the BG-normalization  (Uwano et al 1999, Uwano 2000),
where the one-parameter H{\' e}non-Heiles system was
taken as an example.
\par
Before the outline of the present paper, the inverse problem
of the BG-normalization is explained very briefly. The
BG-normalization has been a powerful method for nonlinear
Hamiltonian systems.
For example, when a two-degree-of-freedom Hamiltonian system
with a 1:1-resonant equilibrium point is given,
the BG-normalization of its Hamiltonian
around the equilibrium point provides an \lq approximate'
Hamiltonian system: The truncation of the normalized Hamiltonian
up to a finite degree is associated with the approximate Hamiltonian
system, which provide a good account of the surface of section
with sufficiently small energies (Kummer 1976, Cushman 1982).
Such a good approximation implies that finding the class
of Hamiltonian systems admitting the same
BG-normalization up to a finite-degree amounts to finding a class
of Hamiltonian systems which admit the surface of section
similar to each other. The following question has been hence posed
by the author as an {\it inverse problem}
of the BG-normalization (Chekanov et al 1998, 2000, Uwano et al 1999):
{\it What kind of polynomial Hamiltonians can be brought into
a given polynomial Hamiltonian in BG-normal form~?}
Since elementary algebraic operations,
differentiation, and integration of polynomials
have to be repeated many times to solve the inverse
problem, computer algebra is worth applying to solve
the inverse problem; see Uwano et al (1999) and Uwano (2000)
for the programme named \lq $\mbox{ANFER}$' and Chekanov et al (1998, 2000) for
\lq $\mbox{GITA}^{-1}$'.
\par
The organization of this paper is outlined as follows.
Section~2 sets up the ordinary and the inverse problems
of the BG-normalization for Hamiltonians, which will be often
referred to as \lq the ordinary problem for Hamiltonians' 
and \lq the inverse problem for BG-normal form Hamiltonians',
respectively, henceforth.
Although the setting-up of the ordinary problem seems to
be merely a review of Moser (1968),
it is of great use to define the ordinary problem in
an mathematically-sound form : Through the review,
the class of canonical transformation to be utilized in the BG-normalization
can be specified explicitly.
In section~3, the one-parameter
H{\' e}non-Heiles Hamiltonian is taken as an example to illustrate
how the ordinary and the inverse problem are proceeded.
From the solution of the inverse problem, it follows that
if the one-parameter H{\' e}non-Heiles system and the perturbed
harmonic oscillator with a homogeneous-quartic polynomial potential
(PHOQP) share the same BG-normal form up to degree-$4$ then both dynamical
systems satisfy the BDIC, so that they are integrable. 
In section~4, the discussion in section~3 made for the one-parameter
H{\' e}non-Heiles Hamiltonian is extended to
the Hamiltonians of the perturbed harmonic oscillators
with homogeneous-cubic polynomial potentials (PHOCP's):
The ordinary and the inverse problems of the BG-normalization
for the PHOCP Hamiltonians are dealt with there.
In the ordinary problem, the Hamiltonians for
PHOCP's are normalized up to degree-$4$.
For the BG-normal form thus obtained, the inverse problem
is solved  up to degree-$4$. Both of the problems are solved
with (a prototype of) the symbolic-computing program ANFER working on
Reduce~3.6 (Uwano 2000).
It is shown that if a PHOCP and a PHOQP share the same BG-normal form
up to degree-$4$ then both oscillators are integrable in the sense that
they satisfy the BDIC.
Section~5 is for concluding remarks
including a conjecture on a further extension.
\section{Setting-up the ordinary and the inverse problems of BG-normalization}
In this section, the ordinary and the inverse problems of the BG-normalization
are reviewed for the two-degree-of-freedom Hamiltonians (Uwano et al 1999).
\subsection{The ordinary problem}
We start with defining the ordinary problem along with a review of
Moser (1968).
Let $\ps$ be the phase space with the canonical coordinates
$(q,p)$ ($q,p \in {\bf R}^2$), and $K(q,p)$ be the Hamiltonian
expressed in power-series form,
\begin{equation}
K(q,p) = \frac{1}{2}\sum_{j=1}^{2} (p_j^2 + q_j^2) +\sum_{k=3}^{\infty}
K_k (q,p),
\label{K}
\end{equation}
where each $K_k (q,p)$ ($k=3,4, \cdots $) is a homogeneous polynomial of
degree-$k$ in $(q,p)$.
\begin{remark}
\label{remark-convergence}
The convergent radius of the power series (\ref{K}) may vanish.
This happens to any $K$ that is not analytic but
differentiable around the origin, for example.
In such a case, the power series (\ref{K}) is considered only in a
formal sense. We will, however, often eliminate the word \lq formal'
from such formal power series henceforth.
\end{remark}
We normalize the Hamiltonian $K$ through the local canonical transformation
of the following form. 
Let $(\xi, \eta)$ be another canonical coordinates working around the origin
of $\ps$. We consider the canonical transformation of $(q,p)$ to
$(\xi , \eta)$ associated with a generating function
(Goldstein 1950) in power-series form,
\begin{equation}
W(q, \eta)=\sum_{j=1}^{2}q_j \eta_j + \sum_{k=3}^{\infty} W_k (q, \eta),
\label{W}
\end{equation}
where each $W_k$ is a homogeneous polynomial of degree-$k$
in $(q, \eta)$ ($k=3,4,\cdots$). The $W(q,\eta)$ is said to be of
the second-type since $W(q,\eta)$ is a function of
the \lq old' position variables $q$ and the \lq new' momentum ones
$\eta$ (see Goldstein 1950).
The canonical transformation associated with $W(q,\eta)$ is given by
the relation
\begin{equation}
\label{can-W}
(q,p) \rightarrow (\xi,\eta)
\quad
\mbox{with}
\quad
p= \frac{\partial W}{\partial q}
\quad
\mbox{and}
\quad
\xi = \frac{\partial W}{\partial \eta},
\end{equation}
which leaves the origin of $\ps$ invariant on account of (\ref{W}).
By $G(\xi,\eta)$, we denote the Hamiltonian brought from $K(q,p)$
by
\begin{equation}
\label{defeq-ord}
G \left(  \frac{\partial W}{\partial \eta}, \eta \right)
=
K \left( q , \frac{\partial W}{\partial q} \right)
\end{equation}
through the transformation (\ref{can-W}).
The BG-normalization for $K$ is accomplished by choosing the generating
function $W(q,\eta)$ in (\ref{defeq-ord}) to put $G(\xi,\eta)$ in BG-normal form:
\begin{definition}
\label{def-ord}
Let $G(\xi,\eta)$ be written in the power-series form,
\begin{equation}
G(\xi , \eta ) = \frac{1}{2}\sum_{j=1}^{2} (\eta_j^2 + \xi_j^2)
+\sum_{k=3}^{\infty} G_k (\xi , \eta) ,
\label{G}
\end{equation}
where each $G_k (\xi , \eta )$ is a
homogeneous polynomial of degree-$k$
($k=3,4,\cdots$) in $(\xi , \eta)$.
The power series $G (\xi , \eta)$ is said to be in BG-normal form
up to degree-$r$ if and only if
\begin{equation}
\left\{ \frac{1}{2}\sum_{j=1}^{2} \left( \eta_j^2 + \xi_j^2 \right) ,
\, G_k (\xi, \eta) \right\}_{\xi , \eta}
=
0 \quad (k=3, \cdots , r)
\label{Poisson}
\end{equation}
holds true, where $\{ \cdot , \cdot \}_{\xi, \eta}$ is the canonical Poisson
bracket  (Arnold 1980) in $(\xi , \eta)$.
\end{definition}
Let us equate the homogeneous-polynomial part
of degree-$k$ ($k=3,4, \cdots $) in (\ref{defeq-ord}).
Then equation (\ref{defeq-ord}) is put into the series of equations,
\begin{equation}
G_{k} (q , \eta ) + (D_{q,\eta} W_{k} ) 
=
K_{k}(q,\eta) + \Phi_k (q,\eta)   \quad (k=3,4, \cdots ),
\label{eq-ord}
\end{equation}
where $D_{q,\eta}$ is the differential operator,
\begin{equation}
\label{def-D}
D_{q,\eta}
=
\sum_{j=1}^{2} \left( 
q_{j} \frac{\partial}{\partial \eta_{j}}
-
\eta_{j} \frac{\partial}{\partial q_{j}}
\right) .
\end{equation}
The $\Phi_k (q, \eta)$ in (\ref{eq-ord})
is the homogeneous polynomial of degree-$k$
in $(q, \eta)$ which is uniquely determined by
$W_{3}, \cdots , W_{k-1}$, $K_{3}, \cdots , K_{k-1}$,  
$G_{3}, \cdots , G_{k-1}$ given:
In particular, we have $\Phi_3 (q,\eta)=0$ and
\begin{equation}
\eqalign{
\Phi_4 (q,\eta)=
\sum_{j=1}^{2} \left (
\frac{1}{2}
\left(
\frac{\partial W_3}{\partial q_j} 
\right)^2
+ 
 \left.
 \frac{\partial K_3}{\partial p_j}
 \right\vert_{(q,\eta)}
\frac{\partial W_3}{\partial q_j}
\right.
\\
\phantom{
\Phi_4 (q,\eta)=\sum_{j=1}^{2} (( \quad
}
\left.
-
\frac{1}{2}
\left(
\frac{\partial W_3}{\partial \eta_j} 
\right)^2
-
 \left.
 \frac{\partial G_3}{\partial \xi_j}
 \right\vert_{(q,\eta)}
\frac{\partial W_3}{\partial \eta_j}
\right).
\label{Phi}
}
\end{equation}
Since we will deal with only the BG-normalization up to
degree-4 in the present paper,
we will not get the expression of
$\Phi_k$ for $k>4$ into more detail
(see Uwano et al 1999, if necessary).
\par
To solve equation (\ref{eq-ord}),
the direct-sum decomposition induced by $D_{q,\eta}$ of the
spaces of homogeneous polynomials is of great use.
Let $V_{k}  (q, \eta)$ denote the vector space
of homogeneous polynomials of degree-$k$ in $(q,\eta)$
with real-valued coefficients ($k = 0,1, \cdots $).
Since the differential operator
$D_{q,\eta}$ acts linearly on each $V_k (q,\eta)$,
the action of $D_{q,\eta}$ naturally induces the direct-sum
decomposition,
\begin{equation}
\label{decomp}
V_{k} (q,\eta) = \image D_{q,\eta}^{(k)}  \oplus \ker D_{q,\eta}^{(k)}
\quad (k=0,1, \cdots),
\end{equation}
of $V_k (q,\eta)$,
where $D_{q,\eta}^{(k)}$ denotes the restriction,
\begin{equation}
\label{restriction}
D_{q,\eta}^{(k)}=\left. D_{q,\eta}\right\vert_{V_{k}(q,\eta)}
\quad (k=3,4,\cdots).
\end{equation}
\begin{remark}
\label{remark-normal}
We have $\ker D_{q,\eta}^{(k)} = \{ 0 \}$ and
$\image D_{q,\eta}^{(k)} = V_k(q,\eta)$ if $k$ is odd. 
\end{remark}
For $D_{q,\eta}$ and $D_{q,\eta}^{(k)}$ ($k=3,4,\cdots$),
we have the following easy to prove.
\begin{lemma}
\label{lemma-Poisson}
Equation (\ref{Poisson}) is equivalent to
\begin{equation}
\label{Lie-deriv}
(D_{q,\eta}(G_k\vert_{(q,\eta)}))(q, \eta)
=
(D_{q,\eta}^{(k)}(G_k\vert_{(q,\eta)}))(q, \eta)
=0
\quad (k=3,\cdots ,r).
\end{equation}
Namely,
$
G_k\vert_{(q,\eta)} \in \ker D_{q,\eta}^{(k)}
$
($k=3, \cdots , r$).
\end{lemma}
According to (\ref{decomp}), let us decompose $K_{k}(q,\eta)$ and $\Phi_k(q,\eta)$
($k=3,4,\cdots$) to be
\begin{equation}
\label{K-Phi-decomp}
\eqalign{
K_k (q,\eta)
=
K_k^{\supim} (q,\eta) + K_k^{\ker} (q,\eta),
\\
\Phi_k (q,\eta)
=
\Phi_k^{\supim} (q,\eta) + \Phi_k^{\ker} (q,\eta),
}
\end{equation}
where
\begin{equation}
\label{K-Phi-component}
\eqalign{
K_k^{\supim} (q,\eta), \Phi_k^{\supim} (q,\eta) \in \image D_{q,\eta}^{(k)},
\\
K_k^{\ker} (q,\eta), \Phi_k^{\ker} (q,\eta) \in \ker D_{q,\eta}^{(k)},
.
}
\end{equation}
Since $G_{k} \in \ker D_{q,\eta}^{(k)}$ by Lemma \ref{lemma-Poisson}
and since $D_{q,\eta}W_{k} \in \image D_{q,\eta}^{(k)}$ ,
we obtain
\begin{equation}
G_{k}(q,\eta)
=
K_k^{\ker} (q,\eta) +\Phi_k^{\ker} (q,\eta) 
\quad
(k=3,4, \cdots),
\label{solution-ord-G}
\end{equation}
as a solution of (\ref{eq-ord}), 
where $W_k$ is chosen to be
\begin{equation}
\label{solution-ord-W}
W_{k}(q,\eta)
=
\left(
\left. D_{q,\eta}^{(k)}\right\vert_{\supim D_{q,\eta}^{(k)}}^{-1}
(K_k^{\supim}\vert_{(q,\eta)}
 +\Phi_k^{\supim}\vert_{(q,\eta)})\right)
(q,\eta).
\end{equation}
What is crucial of (\ref{solution-ord-W})
is that $W_{k} \in \image D_{q,\eta}^{(k)}$
($k=3,4,\cdots$):
For a certain integer $\kappa \geq 3$, let us
consider the sum,
$
{\tilde W}_{\kappa}=
W_{\kappa}+(\mbox{any polynomial in
$\ker D_{q,\eta}^{(\kappa)}$})
$,
where $G_{k}$ and $W_{k}$ with $k < \kappa$ are
given by (\ref{solution-ord-G},\ref{solution-ord-W}).
Even after such a modification, ${\tilde W}_{\kappa}$
satisfy (\ref{eq-ord}) with $k=\kappa$ still,
which leads another series of solutions of (\ref{eq-ord})
with $k > \kappa$. Therefore, under the restriction,
$W_k \in \image D_{q,\eta}^{(k)}$ ($k=3,4,\cdots$),
we can say that
(\ref{solution-ord-G}) with (\ref{solution-ord-W})
is the unique solution of (\ref{defeq-ord}).
To summarize, the ordinary problem is defined
as follows:
\begin{definition}[The ordinary problem]
For a given Hamiltonian $K(q,p)$ in power series (\ref{K}),
bring $K(q,p)$ into the BG-normal form $G(\xi,\eta)$
in power series (\ref{G})
which satisfy (\ref{defeq-ord}) and (\ref{Poisson}) with $r=\infty$,
where the second-type generating function $W$ of the form
(\ref{W}) is chosen to satisfy (\ref{defeq-ord}) and
\begin{equation}
\label{cond-W}
W_{k}(q,\eta) \in \image D_{q,\eta}^{(k)}
\quad (k=3,4,\cdots).
\end{equation}
\end{definition}
\begin{theorem}
\label{theorem-ord}
The BG-normal form $G(\xi,\eta)$ for the Hamiltonian $K(q,p)$
is given by (\ref{G}) with (\ref{solution-ord-G}),
where the second-type generating function $W(q,\eta)$ in power series
(\ref{W}) is chosen to be (\ref{solution-ord-W}).
\end{theorem}
\begin{remark}
\label{remark-convergence-G}
The convergent radius of the BG-normal form $G$ in power-series
(\ref{G}) vanishes in general (see Moser 1968). In such a case,
$G$ is considered only in a formal sense. However, like in
Remark~\ref{remark-convergence} for $K$, we will often
eliminate \lq formal' from such formal power series henceforth.
\end{remark}
\subsection{The inverse problem}
To define the inverse problem of the BG-normalization
appropriately, we review the key equation (\ref{defeq-ord})
of the ordinary problem from a viewpoint of canonical transformations.
Let us regard the power series $W(q,\eta)$ in (\ref{defeq-ord})
as a third-type generating function,
a generating function of the \lq new' position variables $q$
and the \lq old' momentum ones $\eta$ (Goldstein 1950),
which provides the inverse canonical transformation,
\begin{equation}
\label{canonical-xi-eta}
(\xi , \eta ) \rightarrow (q,p)
\quad
\mbox{with}
\quad
\xi = -\frac{\partial (-W)}{\partial \eta}
\quad
\mbox{and}
\quad  
p = -\frac{\partial (-W)}{\partial q},
\end{equation}
of (\ref{can-W}).
Equation (\ref{defeq-ord}) is rewritten as
\begin{equation}
\label{eq-restore}
K( q , -\frac{\partial (-W)}{\partial q} )
=
G ( -\frac{\partial (-W)}{\partial \eta }, \eta) ,
\end{equation}
which is combined with (\ref{canonical-xi-eta})
to show the following.
\begin{lemma}
\label{restore}
Let $G(\xi,\eta)$ of (\ref{G}) be the BG-normal form for the Hamiltonian
$K(q,p)$ of (\ref{K}), which satisfies (\ref{defeq-ord}) with
a second-type generating function $W(q, \eta) \in \image D_{q,\eta}$.
The Hamiltonian $K(q,p)$ is restored from
$G(\xi,\eta)$ through the canonical transformation
(\ref{canonical-xi-eta}) associated with
the third-type generating function
$-W(q, \eta) \in \image D_{q,\eta}$.
\end{lemma}
We are now in a position to pose the inverse problem in the following way: 
Let the Hamiltonian $H(q,p)$ be written in the form,
\begin{equation}
\label{H}
H(q,p)=\frac{1}{2}\sum_{j=1}^{n}
\left( p_j^2 +q_j^2 \right) 
+ \sum_{k=3}^{\infty} H_{k}(q,p),
\end{equation}
where each $H_{k}(q,p)$ ($k=3,4, \cdots $)
is a homogeneous polynomial of degree-$k$
in $(q,p)$.
Further, let a third-type generating function
$S(\eta,q)$ be written in the form,
\begin{equation}
\label{S}
S(q, \eta )
=- \sum_{j=1}^{n} \eta_j q_j 
- \sum_{k=3}^{\infty} S_{k}(q, \eta) ,
\end{equation}
where each $S_{k}(q,\eta)$ ($k=3,4, \cdots $) is
a homogeneous polynomial of degree-$k$
in $(q,\eta)$.
\begin{definition}[The inverse problem]
\label{inv}
For a given BG-normal form, $G(\xi, \eta)$, in power series (\ref{G}),
identify all the Hamiltonians $H(q,p)$ in power series
(\ref{H}) which satisfy
\begin{equation}
\label{defeq-inv}
H( q , -\frac{\partial S}{\partial q} )
=G ( -\frac{\partial S}{\partial \eta }, \eta )
,
\end{equation}
where the third-type generating function $S(q,\eta)$ in power series (\ref{S})
is chosen to satisfy (\ref{defeq-inv}) and
\begin{equation}
\label{cond-S}
S_k(q,\eta) \in \image D_{q,\eta}^{(k)} \quad (k=3,4,\cdots).
\end{equation}
\end{definition}
We solve the inverse problem in the following way.
On equating the homogeneous-polynomial part of
degree-$k$ in (\ref{defeq-inv}),
equation (\ref{defeq-inv}) is put into the series of equations,
\begin{equation}
\label{eq-inv}
H_{k}(q, \eta)-(D_{q,\eta}S_{k})(q, \eta)
= G_{k}(q, \eta) - \Psi_{k}(q, \eta)
\qquad (k=3,4, \cdots),
\end{equation}
where $D_{q, \eta}$ is given by (\ref{def-D}).
The $\Psi_{k} (q,\eta)$ is the homogeneous polynomial
of degree-$k$ in $(q,\eta)$ determined uniquely by
$H_{3}$, $\cdots$, $H_{k-1}$,
$G_{3}$, $\cdots$, $G_{k-1}$, $S_{3}$, $\cdots$,
$S_{k-1}$ given. In particular, we have $\Psi_3 (q,\eta)=0$
and
\begin{equation}
\label{Psi}
\eqalign{
\Psi_4 (q,\eta)
=
\sum_{j=1}^{2} \left (
\frac{1}{2}
\left(
\frac{\partial S_3}{\partial q_j} 
\right)^2
+ 
\left.
\frac{\partial H_3}{\partial p_j}
\right\vert_{(q,\eta)}
\frac{\partial S_3}{\partial q_j}
\right.
\\
\phantom{
\Psi_4 (q,\eta)=
\sum_{j=1}^{2} ((((
}
\left.
-
\frac{1}{2}
\left(
\frac{\partial S_3}{\partial \eta_j} 
\right)^2
-
\left.
\frac{\partial G_3}{\partial \xi_j}
\right\vert_{(q,\eta)}
\frac{\partial S_3}{\partial \eta_j}
\right)
.
}
\end{equation}
\begin{remark}
\label{Phi-Psi}
As is easily seen from (\ref{Phi}) and (\ref{Psi}),
$\Psi_{4}$ takes a similar form to $\Phi_4$ due to (\ref{S}).
Indeed, since the substitution $W=-S$ in (\ref{defeq-ord})
provides (\ref{defeq-inv}),
$\Psi_k$ for each $k>4$ can be obtained as
$\Phi_k$ given by (\ref{eq-ord})
with $H_3, \cdots, H_{k-1}$
and $S_3 , \cdots , S_{k-1}$
in place of
$K_3, \cdots, K_{k-1}$ and $W_3 , \cdots , W_{k-1}$.
Such a similarity will be utilized effectively in future
in writing the program ANFER for the ordinary and the
inverse problems in a unified form.
\end{remark}
Like in the ordinary problem, we solve (\ref{eq-inv}) for
$H_{k}$ and $S_{k}$ by using the direct-sum decomposition (\ref{decomp})
of $V_k (q,\eta)$, the vector spaces of homogeneous
polynomials of degree-$k$ ($k=3,4,\cdots$). Let us decompose
$H_k$ and $\Psi_k$ to be
\begin{equation}
\label{H-Psi-decomp}
\eqalign{
H_k (q,\eta) = H_k^{\supim}(q,\eta) + H_k^{\ker} (q,\eta),
\\
\Psi_k (q,\eta) = \Psi_k^{\supim} (q,\eta) + \Psi_k^{\ker} (q,\eta),
}
\end{equation}
where
\begin{equation}
\label{H-Psi-component}
\eqalign{
K_k^{\supim} (q,\eta), \Psi_k^{\supim} (q,\eta) \in \image D_{q,\eta}^{(k)},
\\
K_k^{\ker} (q,\eta), \Psi_k^{\ker} (q,\eta) \in \ker D_{q,\eta}^{(k)},
}
\end{equation}
Then on equating $\ker D_{q,\eta}^{(k)}$-part in
(\ref{eq-inv}),
$H_{k}^{\ker}$ is determined to be
\begin{equation}
\label{solution-inv-ker}
H_{k}^{\ker}(q, \eta) 
= G_{k}(q, \eta) - \Psi_{k}^{\ker}(q, \eta).
\end{equation}
Equating $\image D_{q,\eta}$-part of (\ref{eq-inv}),
we have 
\begin{equation}
\label{eq-inv-image}
H_{k}^{\supim}(q, \eta) 
-\left( D_{q, \eta} S_{k} \right)(q, \eta)
= - \Psi_{k}^{\supim}(q, \eta) .
\end{equation}
Since the pair of unidentified polynomials,
$H_{k}^{\supim}$ and $S_{k}$ exists in (\ref{eq-inv-image}),
$H_{k}^{\supim}$ is not determined uniquely in contrast with
$H_{k}^{\ker}$;
such a non-uniqueness is of the very nature of the inverse problem.
Accordingly, equation (\ref{eq-inv-image}) is solved to as
\begin{eqnarray}
\label{solution-inv-image-H}
H_{k}^{\supim}(q,\eta) \in \image D_{q,\eta}^{(k)}:
\; \mbox{chosen arbitrarily},
\\
\label{solution-inv-image-S}
S_{k}(q,\eta)
=
\left(
\left. D_{q,\eta}^{(k)}\right\vert_{\supim D_{q,\eta}^{(k)}}^{-1}
(H_k^{\supim}\vert_{(q,\eta)} +\Psi_k^{\supim}\vert_{(q,\eta)})
\right)(q,\eta),
\end{eqnarray}
($k=3,4,\cdots$). Now we have the following.
\begin{theorem}
\label{theorem-inv}
For a given BG-normal form $G(\xi,\eta)$ in power series (\ref{G}),
the solution $H(q,p)$ of the inverse problem is given by
(\ref{H}) subject to (\ref{solution-inv-ker}) and
(\ref{solution-inv-image-H}),
where the third-type generating function $S(q,\eta)$ in
(\ref{defeq-inv}) is chosen to be (\ref{S}) subject to
(\ref{solution-inv-image-S}).
\end{theorem}
\subsection{The degree-$2\delta$ ordinary and inverse problems}
In the preceding subsections, we have defined the ordinary and the
inverse problems of BG-normalization, and then find their
solutions in power-series.
From a practical point of view, however, we usually deal with
the BG-normal forms not in power-series but in polynomial.
Indeed, as mentioned in section~1, when we utilize
the BG-normalization to provide an approximate system
for a given system, we truncate the normalized Hamiltonian
up to a finite degree. Hence it is natural to think of
a \lq finite-degree version' of both the ordinary and the inverse
problems (Uwano et al 1999).
\begin{definition}
[The degree-$2\delta$ ordinary problem]
\label{deg-2d-ord}
For a given Hamiltonian $K(q,p)$ of the form (\ref{K})
(possibly in polynomial form) and an integer $\delta \geq 2$,
bring $K$ into the polynomial $G(\xi,\eta)$ of degree-$2\delta$
in BG-normal form which satisfy (\ref{defeq-ord}) up to degree-$2\delta$,
where the second-type generating function $W(q,\eta)$ in (\ref{defeq-ord})
is chosen to be the polynomial of degree-$2\delta$ subject to
(\ref{defeq-ord}) up to degree-$2\delta$ and (\ref{cond-W})
with $k=3,\cdots,2\delta$.
\end{definition}
\begin{remark}
\label{reason-degree}
The reason why we think of only the even-($2\delta$-)degree case 
is that the BG-normal form of any 1:1 resonant Hamiltonian
consists of even-degree terms only (see Remark~\ref{remark-normal} and
Lemma~\ref{lemma-Poisson}).
\end{remark}
\begin{definition}[The degree-$2\delta$ inverse problem]
\label{deg-2d-inv}
For a given BG-normal form, $G(\xi, \eta)$, of degree-$2\delta$
with an integer $\delta \geq 2$, identify all the
polynomial-Hamiltonians $H(q,p)$ of degree-$2\delta$
which satisfy (\ref{defeq-inv}) up to degree-$2\delta$,
where the third-type generating function $S(q, \eta)$
is chosen to be the polynomial of degree-$2\delta$
subject to (\ref{defeq-inv}) up to degree-$2\delta$ and
(\ref{cond-S}) with $k=3,\cdots,2\delta$.
\end{definition}
In closing this section, we wish to mention of the way to
solve the (degree-$2\delta$) ordinary and the inverse problems
on computer algebra.  Although the discussion
throughout this section is mathematically complete,
it is not easy to calculate even on computer algebra
(\ref{solution-ord-G},\ref{solution-ord-W}) and
(\ref{solution-inv-ker},\ref{solution-inv-image-H},\ref{solution-inv-image-S})
as they present because we are faced with a highly combinatorial
difficulty in calculating $\Phi_{k}$ and $\Psi_{k}$:
To calculate $\Psi_{k}$ for example, $G_{3}, \cdots , G_{k-1}$,
$S_{3},\cdots, S_{k-1}$, and $H_{3} , \cdots, H_{k-1}$
have to be kept on computer. What is worse is that
the dimension,
$
\sum_{h=1}^{4} {4 \choose h}  {\ell-1 \choose h-1}  ,
$
of $V_{\ell}$ to which $G_{\ell}$, $S_{\ell}$ and $H_{\ell}$
belong rises in a combinatorial manner as $\ell$ increases,
where the symbol $ {\cdot \choose \cdot} $ indicates
the binomial coefficient. These facts will cause a combinatorial
increase of the memory-size on computer required for calculation.
To get rid of such a difficulty, 
we break the transformations, (\ref{defeq-ord}) and (\ref{defeq-inv})
into a recursion of certain canonical transformations of simpler form,
which will be presented in Appendix for the degree-$4$ case.
\section{Example: The one-parameter H{\' e}non-Heiles system}
In this section, we take the one-parameter H{\' e}non-Heiles-Hamiltonian
\begin{equation}
\label{K-HH}
K_{\mu}(q,p)
=
\frac{1}{2}\sum_{j=1}^{2} (p_j^2 + q_j^2)
+q_1^2q_2 +\mu q_2^3  \qquad (\mu \in {\bf R}),
\end{equation}
as an example to illustrate how the ordinary and the inverse problems
are proceeded. As is mentioned in section~1, this example inspired
the present work.
\subsection{The degree-$4$ ordinary and inverse problems}
In order to present the results in compact forms, the complex variables
defined by 
\begin{equation}
\label{z-zeta}
z_j = q_j + i p_j , \qquad
\zeta_j=\xi_j + i \eta_j
\quad
(j=1,2),
\end{equation}
will be of great use.
In terms of $z$ and ${\ovl z}$,
$K_{\mu}$ is written in the form,
\begin{equation}
\label{cplx-K-HH}
\eqalign{
K_{\mu}(q,p)
=
\frac{1}{2}(z_1{\ovl z}_1 + z_2 {\ovl z}_2)
+\frac{\mu}{8}(
z_2^3 + 3z_2^2 {\ovl z}_2 + 3 z_2 {\ovl z}_2^2 +{\ovl z}_2^3
)
\\
\phantom{K_{\mu}(q,p)=}
+\frac{1}{8}(
z_1^2z_2+z_1^2{\ovl z}_2 +{\ovl z}_1 z_2^2 + {\ovl z}_1 {\ovl z}_2^2
+2 z_1 {\ovl z}_1 z_2 +2 z_1 {\ovl z}_1 {\ovl z}_2 
).
}
\end{equation}
Using a prototype of ANFER (Uwano et al 1999, Uwano 2000),
we see that the BG-normal form for $K_{\mu}$
is given, up to degree-$4$, by,
\begin{equation}
\eqalign{
G_{\mu}(\xi , \eta)
=
\frac{1}{2}(\zeta_1{\ovl \zeta}_1 + \zeta_2 {\ovl \zeta}_2)
\\
\phantom{G_{\mu}(\xi , \eta)=}
+
\frac{1}{48}
\left\{
-5\zeta_1^2{\ovl \zeta}_1^2
-45\mu^2\zeta_2^2{\ovl \zeta}_2^2
-(8+36\mu)\zeta_1\zeta_2{\ovl \zeta}_1{\ovl \zeta}_2
\right.
\\
\phantom{G_{\mu}(\xi , \eta)======}
\left.
+3\mu\zeta_1^2{\ovl \zeta}_2^2
+3\mu\zeta_2^2{\ovl \zeta}_1^2
-6\zeta_1^2{\ovl \zeta}_2^2
-6\zeta_2^2{\ovl \zeta}_1^2
\right\},
}
\end{equation}
which is well known (see Kummer 1976, Cushman 1982).
\indent
We proceed to the degree-$4$ inverse problem for the BG-normal form
$G_{\mu}$ in turn.
Solving (\ref{defeq-inv}) with $G_{\mu}$ in place of $G$
by ANFER, we have the Hamiltonians in polynomial
of degree-$4$ of the following form as the solution;
\begin{equation}
\label{H-HH}
H_{\mu}(q,p)
=
\frac{1}{2}\sum_{j=1}^{2}(p_j^2 + q_j^2)
+H_{\mu,3}(q,p) + H_{\mu,4}(q,p)
\end{equation}
with
\begin{eqnarray}
H_{\mu,3}(q,p)
\nonumber
\\
=
a_1z_1^3+a_2z_1^2z_2+a_3z_1z_2^2+a_4z_2^3+a_5z_1^2{\ovl z}_1
\nonumber
\\
\phantom{=}
+a_6z_1^2{\ovl z}_2+a_7z_1z_2{\ovl z}_1
+a_8z_1z_2{\ovl z}_2+a_9z_2^2{\ovl z}_1
+a_{10}z_2^2{\ovl z}_2
\label{H3-HH}
\\
\phantom{=}
+{\ovl a}_1{\ovl z}_1^3
+{\ovl a}_2{\ovl z}_1^2{\ovl z}_2+{\ovl a}_3{\ovl z}_1{\ovl z}_2^2
+{\ovl a}_4{\ovl z}_2^3+{\ovl a}_5z_1{\ovl z}_1^2
\nonumber
\\
\phantom{=}
+{\ovl a}_6z_2{\ovl z}_1^2
+{\ovl a}_7z_1{\ovl z}_1{\ovl z}_2
+{\ovl a}_8z_2{\ovl z}_1{\ovl z}_2
+{\ovl a}_9z_1{\ovl z}_2^2
+{\ovl a}_{10}z_2{\ovl z}_2^2
\nonumber
\end{eqnarray}
and
\begin{eqnarray}
\nonumber
H_{\mu,4}(q,p)
\\ \nonumber
=c_1z_1^4+c_2z_1^3z_2+c_3z_1^2z_2^2+c_4z_1z_2^3+c_5z_2^4
+c_6z_1^3{\ovl z}_1+c_7z_1^3{\ovl z}_2
\\ \nonumber
\phantom{=}
+c_8z_1^2z_2{\ovl z}_1+c_9z_1^2z_2{\ovl z}_2+c_{10}z_1z_2^2{\ovl z}_1
+c_{11}z_1z_2^2{\ovl z}_2+c_{12}z_2^3{\ovl z}_1+c_{13}z_2^3{\ovl z}_2
\\ \nonumber 
\phantom{=}
+{\ovl c}_1{\ovl z}_1^4+{\ovl c}_2{\ovl z}_1^3{\ovl z}_2
+{\ovl c}_3{\ovl z}_1^2{\ovl z}_2^2+{\ovl c}_4{\ovl z}_1{\ovl z}_2^3
+{\ovl c}_5{\ovl z}_2^4+{\ovl c}_6z_1{\ovl z}_1^3+{\ovl c}_7z_2{\ovl z}_1^3
\\ \nonumber
\phantom{=}
+{\ovl c}_8z_1{\ovl z}_1^2{\ovl z}_2
+{\ovl c}_9z_2{\ovl z}_1^2{\ovl z}_2
+{\ovl c}_{10}z_1{\ovl z}_1{\ovl z}_2^2
+{\ovl c}_{11}z_2{\ovl z}_1{\ovl z}_2^2
+{\ovl c}_{12}z_1{\ovl z}_2^3+{\ovl c}_{13}z_2{\ovl z}_2^3
\\ \nonumber
\phantom{=}
+8 \left(
a_9{\ovl a}_{10}z_2^2{\ovl z}_1{\ovl z}_2
+a_9{\ovl a}_9z_1z_2{\ovl z}_1{\ovl z}_2
+a_{10}{\ovl a}_9z_1z_2{\ovl z}_2^2
\right.
\\
\nonumber
\phantom{====}
\left.
+a_5{\ovl a}_6z_1z_2{\ovl z}_1^2
+a_6{\ovl a}_5z_1^2{\ovl z}_1{\ovl z}_2
+a_6{\ovl a}_6z_1z_2{\ovl z}_1{\ovl z}_2
\right)
\\ \nonumber
\phantom{=}
+6 \left(
a_1{\ovl a}_1z_1^2{\ovl z}_1^2
+a_{10}{\ovl a}_{10}z_2^2{\ovl z}_2^2
+a_4{\ovl a}_4z_2^2{\ovl z}_2^2
+a_5{\ovl a}_5z_1^2{\ovl z}_1^2
\right)
\\ \nonumber
\phantom{=}
+4\left(
a_8{\ovl a}_5z_1z_2{\ovl z}_1{\ovl z}_2
+a_8{\ovl a}_6z_2^2{\ovl z}_1{\ovl z}_2
+a_8{\ovl a}_9z_1^2{\ovl z}_2^2
+a_9{\ovl a}_7z_1z_2{\ovl z}_1^2
\right.
\\ \nonumber
\phantom{====}
+a_9{\ovl a}_8z_2^2{\ovl z}_1^2
+a_1{\ovl a}_2z_1^2{\ovl z}_1{\ovl z}_2
+a_{10}{\ovl a}_7z_1z_2{\ovl z}_1{\ovl z}_2
+a_2{\ovl a}_1z_1z_2{\ovl z}_1^2
\\ 
\phantom{====}
+a_3{\ovl a}_4z_1z_2{\ovl z}_2^2
+a_4{\ovl a}_3z_2^2{\ovl z}_1{\ovl z}_2
+a_5{\ovl a}_8z_1z_2{\ovl z}_1{\ovl z}_2
+a_6{\ovl a}_7z_1^2{\ovl z}_2^2
\label{H4-HH}
\\ \nonumber
\phantom{====}
\left.
+a_6{\ovl a}_8z_1z_2{\ovl z}_2^2
+a_7{\ovl a}_{10}z_1z_2{\ovl z}_1{\ovl z}_2
+a_7{\ovl a}_6z_2^2{\ovl z}_1^2
+a_7{\ovl a}_9z_1^2{\ovl z}_1{\ovl z}_2
\right)
\\ \nonumber
\phantom{=}
+\frac{8}{3}\left(
a_2{\ovl a}_2z_1z_2{\ovl z}_1{\ovl z}_2
+a_3{\ovl a}_3z_1z_2{\ovl z}_1{\ovl z}_2
\right)
\\ \nonumber
\phantom{=}
+2\left(
a_8{\ovl a}_{10}z_1z_2{\ovl z}_2^2
+a_8{\ovl a}_7z_1^2{\ovl z}_1{\ovl z}_2
+a_8{\ovl a}_7z_1z_2{\ovl z}_2^2
+a_8{\ovl a}_8z_2^2{\ovl z}_2^2
\right.
\\ \nonumber
\phantom{====}
+a_1{\ovl a}_3z_1^2{\ovl z}_2^2
+a_{10}{\ovl a}_8z_2^2{\ovl z}_1{\ovl z}_2
+a_2{\ovl a}_4z_1^2{\ovl z}_2^2
+a_3{\ovl a}_1z_2^2{\ovl z}_1^2
\\ \nonumber
\phantom{====}
+a_4{\ovl a}_2z_2^2{\ovl z}_1^2
+a_5{\ovl a}_7z_1^2{\ovl z}_1{\ovl z}_2
+a_7{\ovl a}_5z_1z_2{\ovl z}_1^2
+a_7{\ovl a}_7z_1^2{\ovl z}_1^2
\\ \nonumber
\phantom{====}
+a_7{\ovl a}_8z_1z_2{\ovl z}_1^2
+a_7{\ovl a}_8z_2^2{\ovl z}_1{\ovl z}_2
-a_8{\ovl a}_6z_1z_2{\ovl z}_1^2
-a_7{\ovl a}_9z_1z_2{\ovl z}_2^2
\\ \nonumber
\phantom{====}
-a_9{\ovl a}_5z_2^2{\ovl z}_1^2
-a_9{\ovl a}_7z_2^2{\ovl z}_1{\ovl z}_2
-a_9{\ovl a}_9z_2^2{\ovl z}_2^2
-a_{10}{\ovl a}_6z_2^2{\ovl z}_1^2
\\ \nonumber
\phantom{====}
\left.
-a_5{\ovl a}_9z_1^2{\ovl z}_2^2
-a_6{\ovl a}_{10}z_1^2{\ovl z}_2^2
-a_6{\ovl a}_6z_1^2{\ovl z}_1^2
-a_6{\ovl a}_8z_1^2{\ovl z}_1{\ovl z}_2
\right)
\\ \nonumber
\phantom{=}
+\frac{4}{3}\left(
+a_2{\ovl a}_3z_1^2{\ovl z}_1{\ovl z}_2+a_2{\ovl a}_3z_1z_2{\ovl z}_2^2
+a_3{\ovl a}_2z_1z_2{\ovl z}_1^2+a_3{\ovl a}_2z_2^2{\ovl z}_1{\ovl z}_2
\right)
\\ \nonumber
\phantom{=}
+\frac{2}{3}\left(
a_2{\ovl a}_2z_1^2{\ovl z}_1^2
+a_3{\ovl a}_3z_2^2{\ovl z}_2^2
\right)
\\ \nonumber
\phantom{=}
+\frac{1}{48}\left(
-8z_1z_2{\ovl z}_1{\ovl z}_2-5z_1^2{\ovl z}_1^2
-6z_1^2{\ovl z}_2^2-6z_2^2{\ovl z}_1^2
\right.
\\ \nonumber
\phantom{=====}
\left.
-36\mu z_1z_2{\ovl z}_1{\ovl z}_2-45\mu^2z_2^2{\ovl z}_2^2
+3\mu z_1^2{\ovl z}_2^2+3\mu z_2^2{\ovl z}_1^2
\right),
\end{eqnarray}
where $a_h$ ($h=1,\cdots,10$) and $c_{\ell}$ ($\ell =1, \cdots ,13$)
are complex-valued parameters chosen arbitrarily.
In (\ref{H4-HH}), the polynomial with the coefficients $(c_{\ell})$
expresses $H_{\mu,4}^{\supim}$ and the polynomial whose coefficients
are written in terms of $(a_h)$ and $(f_k)$ does $H_{\mu,4}^{\ker}$.
From the lengthy expression (\ref{H-HH}, \ref{H3-HH}, \ref{H4-HH}),
one might understand an effectiveness of
computer algebra in the inverse problem.
\par
It is worth pointing out that if we choose $(a_h)$ and
$(c_{\ell})$ to be
\begin{equation}
\label{special1-HH}
\eqalign{
a_1=a_3=a_5=a_8=a_9=0,
\\
2a_2=2a_6=a_7=\frac{1}{4}, \; \; 
3a_4=a_{10}=\frac{3\mu}{8},
}
\end{equation}
and
\begin{equation}
\label{special2-HH}
c_{\ell}=0 \quad  (\ell=1,\cdots,13),
\end{equation}
respectively, $H_{\mu}$ becomes equal to
the one-parameter H{\' e}non-Heiles Hamiltonian $K_{\mu}$.
\subsection{The Bertrand-Darboux integrability condition}
We wish to find the condition for $(a_h)$ and $(c_{\ell})$
to bring $H_{\mu}$ into the Hamiltonian of 
the perturbed harmonic oscillator with a homogeneous-quartic polynomial
potential (PHOQP).
To bring $H_{\mu}$ into a PHOQP-Hamiltonian, we have to
make $H_{\mu,3}$ vanish. The vanishment of $H_{\mu,3}$ is realized
by the substitution
\begin{equation}
\label{vanish-H3}
a_h = 0 \quad (h=1, \cdots , 10)
\end{equation}
in (\ref{H3-HH}).
We bring $H_{\mu , 4}\vert_{a=0}(q,p)$ into a homogeneous
polynomial of degree-$4$ in $q$ in turn.
To do this,
we have to find the set of non-vanishing $((c_{\ell}), (\lambda_m))$
for which the following identities for $z$ hold true;
\begin{eqnarray}
\lambda_1 q_1^4
=
c_1z_1^4+c_6z_1^3{\ovl z}_1-\frac{5}{48}z_1^2{\ovl z}_1^2
+{\ovl c}_1{\ovl z}_1^4+{\ovl c}_6z_1{\ovl z}_1^3
,
\nonumber \\
\lambda_2 q_1^3q_2 
=
c_2z_1^3z_2++c_7z_1^3{\ovl z}_2+c_8z_1^2z_2{\ovl z}_1
\\
\phantom{\lambda_2 q_1^3q_2===}
+{\ovl c}_2{\ovl z}_1^3{\ovl z}_2
+{\ovl c}_7z_2{\ovl z}_1^3
+{\ovl c}_8z_1{\ovl z}_1^2{\ovl z}_2
,
\nonumber \\
\lambda_3 q_1 q_2^3
=
c_{11}z_1z_2^2{\ovl z}_2+c_{12}z_2^3{\ovl z}_1+c_4z_1z_2^3
\nonumber \\
\phantom{\lambda_3 q_1 q_2^3===}
+{\ovl c}_4{\ovl z}_1{\ovl z}_2^3
+{\ovl c}_{12}z_1{\ovl z}_2^3
+{\ovl c}_{11}z_2{\ovl z}_1{\ovl z}_2^2
,
\nonumber \\
\lambda_4 q_2^4
=
c_5z_2^4+c_{13}z_2^3{\ovl z}_2
-\frac{15}{16}\mu^2z_2^2{\ovl z}_2^2
+{\ovl c}_{13}z_2{\ovl z}_2^3
+{\ovl c}_5{\ovl z}_2^4
,
\label{identity-HH}
\\
\lambda_5 q_1^2 q_2^2
=
c_{10}z_1z_2^2{\ovl z}_1+c_3z_1^2z_2^2+c_9z_1^2z_2{\ovl z}_2
\nonumber \\
\phantom{\lambda_5 q_1^2 q_2^2===}
+{\ovl c}_{10}z_1{\ovl z}_1{\ovl z}_2^2
+{\ovl c}_3{\ovl z}_1^2{\ovl z}_2^2
+{\ovl c}_9z_2{\ovl z}_1^2{\ovl z}_2
\nonumber \\
\phantom{\lambda_5 q_1^2 q_2^2====}
+\frac{\mu}{16} z_1^2{\ovl z}_2^2+\frac{\mu}{16} z_2^2{\ovl z}_1^2
-\frac{3\mu}{4} z_1z_2{\ovl z}_1{\ovl z}_2
\nonumber \\
\phantom{\lambda_5 q_1^2 q_2^2=====}
-\frac{1}{6}z_1z_2{\ovl z}_1{\ovl z}_2
-\frac{1}{8}z_1^2{\ovl z}_2^2-\frac{1}{8}z_2^2{\ovl z}_1^2
,
\nonumber
\end{eqnarray}
where $\lambda_m$ ($m=1,\cdots , 5$) are real-valued parameters,
and $q_j=(z_j+{\ovl z}_j)/2$ ($j=1,2$).
To make the identities hold true from the first to the fourth,
we have to choose the parameters appearing in those identities
to be
\begin{equation}
\label{cond-c-l-1}
\eqalign{
\lambda_1=16c_1=4c_6=-\frac{5}{18},
\quad
\lambda_2=c_2=c_7=c_8=0,
\\ 
\lambda_3=c_4=c_{11}=c_{12}=0,
\quad
\lambda_4=16c_5=4c_{13}=-\frac{5\mu^2}{2}.
}
\end{equation}
The fifth identity holds true if and only if
the overdetermined system of equations,
\begin{equation}
\label{eq-c-l-2}
\eqalign{
\lambda_5=16c_3=8c_9=8c_{10},
\\
\lambda_5=- 3\mu - \frac{2}{3},
\\
\lambda_5= \mu - 2,
}
\end{equation}
admits a solution. By a simple calculation,
we see that (\ref{eq-c-l-2}) admits the solution,
\begin{equation}
\label{cond-c-l-2}
\lambda_5=16c_3=8c_9=8c_{10}=-\frac{5}{3},
\end{equation}
if and only if $\mu$ satisfies 
\begin{equation}
\label{cond-mu}
\mu=\frac{1}{3}.
\end{equation}
To summarize, we have the following.
\begin{theorem}
\label{PHOQP-HH}
The one-parameter H{\' e}non-Heiles Hamiltonian
$K_{\mu}(q,p)$ shares its BG-normal form up to degree-$4$
with the Hamiltonian of the perturbed
harmonic oscillator with a homogeneous-quartic
polynomial potential (PHOQP)
if and only if the parameter $\mu$ satisfies (\ref{cond-mu}).
Under (\ref{cond-mu}), the PHOQP-Hamiltonian sharing the BG-normal
form with $K_{1/3}$ is given by
\begin{equation}
\label{Q-HH}
Q(q,p)
=
\frac{1}{2}\sum_{j=1}^{2}(p_j^2 + q_j^2) 
-\frac{5}{18}(q_1^4 +6q_1^2q_2^2 +q_2^4).
\end{equation}
\end{theorem}
After theorem \ref{PHOQP-HH}, one might ask whether
$K_{1/3}$ and $Q$ have specific meanings or not.
Those who are familiar with the separability of dynamical systems
may recognize immediately that $H_{1/3}$ and $Q$ are separable
in $q_1 \pm q_2$ (see Perelomov 1990).
We can hence answer this question affirmatively owing to the Bertrand-Darboux
theorem concerning not only the separability but also
the integrability. The theorem is stated as follows
(see Marshall and Wojciechowski 1988,
Yamaguchi and Nambu 1998, Grosche et al 1995, for example).
\begin{theorem}[Bertrand-Darboux]
\label{BDIC-theorem}
Let $F$ be a natural Hamiltonian of the form,
\begin{equation}
\label{F}
F(q,p)
=
\frac{1}{2}\sum_{j=1}^{2}p_j^2 +V(q),
\end{equation}
where $V(q)$ a differentiable function in $q$.
Then, the following three statements are equivalent for the Hamiltonian
system with $F$. 
\begin{description}
\item[{(1)}]
There exists a set of non-vanishing real-valued constants,
$(\alpha, \beta, \beta^{\prime}, \gamma, \gamma^{\prime})$,
for which $V(q)$ satisfies
\begin{equation}
\label{BDIC-general}
\eqalign{
\left(
\frac{\partial^2 V}{\partial q_2^2}
-\frac{\partial^2 V}{\partial q_1^2}
\right)
(-2\alpha q_1q_2-\beta^{\prime}q_2 -\beta q_1 + \gamma)
\\
\phantom{xxxx}
+2 \frac{\partial^2 V}{\partial q_1 \partial q_2}
(\alpha q_2^2 -\gamma q_1^2 +\beta q_2 -\beta^{\prime}q_1
+\gamma^{\prime})
\\
\phantom{xxxxxxxx}
+\frac{\partial V}{\partial q_1}
(6\alpha q_2 + 3\beta)
-
\frac{\partial V}{\partial q_2}
(6\alpha q_1 + 3\beta^{\prime})=0 .
}
\end{equation}
\item[{(2)}]
The Hamiltonian system with $F$ admits an integral of motion
quadratic in momenta.
\item[{(3)}]
The Hamiltonian $F$ is separable in either Cartesian, polar,
parabolic or  elliptic coordinates.
\end{description}
\end{theorem}
\indent
Due to the statement (2) in theorem~\ref{BDIC-theorem}, a natural Hamiltonian system
with $F$ is always integrable if (\ref{BDIC-general}) holds true.
In this regard, we will refer to (\ref{BDIC-general}) as the Bertrand-Darboux
integrability condition (BDIC) henceforth.
\par
For the PHOCP's and the PHOQP's, Yamaguchi and Nambu (1998) have given
a more explicit expression of the BDIC (\ref{BDIC-general}) convenient
for our purpose:
\begin{lemma}
\label{lemma-YN}
Let ${\cal F}^{(k)}(q,p)$ ($k=3,4$) be the Hamiltonians of the form,
\begin{equation}
\label{F3-F4}
{\cal F}^{(k)}(q,p)= \frac{1}{2}\sum_{j=1}^{2}(p_j^2+q_j^2 )
+{\cal P}^{(k)}(q) \qquad (k=3,4),
\end{equation}
with 
\begin{equation}
\label{P3}
{\cal P}^{(3)}(q)
=
f_1q_1^3 + f_2 q_1^2 q_2 + f_3 q_1 q_2^2 + f_4 q_2^3,
\end{equation}
or
\begin{equation}
\label{P4}
{\cal P}^{(4)}(q)
=
g_1q_1^4 + g_2 q_1^3 q_2 + g_3 q_1^2 q_2^2
+ g_4 q_1 q_2^3 + g_5 q_2^4,
\end{equation}
where $f_k$ ($k=1, \cdots ,4$) and
$g_{\ell}$ ($\ell=1,\cdots , 5$) are real-valued
parameters.
For the PHOCP with ${\cal F}^{(3)}(q,p)$,
the BDIC (\ref{BDIC-general}) is equivalent to either of
the following conditions, (\ref{BDIC-P3-a}), (\ref{BDIC-P3-b}),
or (\ref{BDIC-P3-c});
\numparts
\begin{eqnarray}
3(f_1f_3+f_2f_4)-(f_2^2+f_3^2)=0,
\label{BDIC-P3-a}
\\
f_1=2f_3, \quad f_2=f_4=0,
\label{BDIC-P3-b}
\\
f_4=2f_2, \quad f_1=f_3=0.
\label{BDIC-P3-c}
\end{eqnarray}
\endnumparts
For the PHOQP with ${\cal F}^{(4)}(q,p)$,
the BDIC (\ref{BDIC-general}) is equivalent to either of the following
conditions, (\ref{BDIC-P4-a}) or (\ref{BDIC-P4-b});
\numparts
\begin{eqnarray}
\left\{
\begin{array}{l}
g_3=2g_1=2g_5,
\\
g_2=g_4=0,
\end{array}
\right. 
\label{BDIC-P4-a}
\\
\left\{
\begin{array}{l}
9g_2^2+4g_3^2-24g_1g_3-9g_2g_4=0,
\\ \noalign{\vskip4pt}
9g_4^2+4g_3^2-24g_3g_5-9g_2g_4=0,
\\ \noalign{\vskip4pt}
(g_2+g_4)g_3-6(g_1g_4+g_2g_5)=0.
\end{array}
\right.
\label{BDIC-P4-b}
\end{eqnarray}
\endnumparts
\end{lemma}
\indent
The integrability of the H{\' e}non-Heiles system with
$\mu=1/3$ and the PHOQP with $Q$ can be confirmed now by showing
that (\ref{BDIC-general}) holds true both for $K_{1/3}$ and
$Q$. Let us start with showing the integrability of the
one-parameter H{\' e}non-Heiles system with $\mu =1/3$. If we choose $(f_h)$ to be
\begin{equation}
\label{P3-HH1/3}
f_1=f_3=0, \quad f_2=1, \quad f_4=\frac{1}{3},
\end{equation}
the Hamiltonian ${\cal F}^{(3)}$ becomes $K_{1/3}$.
Evidently, the $(f_h)$ given by (\ref{P3-HH1/3}) satisfy
the BDIC (\ref{BDIC-P3-a}), so that the one-parameter H{\' e}non-Heiles system
with $\mu=1/3$ is integrable.
\par
We proceed to showing the integrability of the PHOQP with $Q$
given by (\ref{Q-HH}) in turn.
In order to bring ${\cal F}^{(4)}$ into $Q$,
we choose $(g_{\ell})$ to be
\begin{equation}
\label{P4-HH}
6g_1=g_3=6g_5=-\frac{5}{3},
\quad
g_2=g_4=0.
\end{equation}
By a simple calculation, we see that
the $(g_{\ell})$ given by (\ref{P4-HH})
satisfy the BDIC (\ref{BDIC-P4-b}),
so that the PHOQP with $Q$ is integrable.
To summarize, we have the following.
\begin{theorem}
\label{theorem-BDIC-HH}
If the one-parameter H{\' e}non-Heiles system
and the perturbed harmonic oscillator with a
homogeneous-quartic polynomial potential
admit the same BG-normalization up to degree-$4$,
then both dynamical systems are integrable in the sense that
they satisfy the Bertrand-Darboux integrability
condition.
\end{theorem}
\section{Extension: The perturbed harmonic oscillators
with homogeneous-cubic polynomial potentials}
\subsection{The degree-$4$ ordinary and the inverse problems of PHOCP}
In this section, the discussion in section~3 for the one-parameter
H{\' e}non-Heiles system is extended to
the perturbed harmonic oscillators
with homogeneous-cubic polynomial potentials (PHOCP's).
To start with, we solve the degree-$4$ ordinary problem
for the PHOCP-Hamiltonian ${\cal F}^{(3)}(q,p)$ given by
(\ref{F3-F4}) with (\ref{P3}).
By ANFER, we see that the BG-normal form for ${\cal F}^{(3)}$
is given, up to degree-$4$, by
\begin{eqnarray}
{\cal G}(\xi ,\eta)
=\frac{1}{2}\left( \zeta_1{\ovl \zeta}_1 + \zeta_2{\ovl \zeta}_2
\right)
\nonumber
\\
\phantom{{\cal G}(\xi ,\eta)x}
-\frac{15}{16}
(f_1^2\zeta_1^2{\ovl \zeta}_1^2+f_4^2\zeta_2^2{\ovl \zeta}_2^2)
-\frac{3}{4}(f_1f_3\zeta_1\zeta_2{\ovl \zeta}_1{\ovl \zeta}_2+f_2f_4\zeta_1\zeta_2{\ovl \zeta}_1{\ovl \zeta}_2)
\nonumber
\\
\phantom{{\cal G}(\xi ,\eta)x}
-\frac{5}{8}(f_1f_2\zeta_1^2{\ovl \zeta}_1{\ovl \zeta}_2
+f_1f_2\zeta_1\zeta_2{\ovl \zeta}_1^2
+f_3f_4\zeta_1\zeta_2{\ovl \zeta}_2^2
+f_3f_4\zeta_2^2{\ovl \zeta}_1{\ovl \zeta}_2)
\nonumber
\\
\phantom{{\cal G}(\xi ,\eta)x}
-\frac{5}{24}(f_2f_3\zeta_1^2{\ovl \zeta}_1{\ovl \zeta}_2
+f_2f_3\zeta_1\zeta_2{\ovl \zeta}_1^2
+f_2f_3\zeta_1\zeta_2{\ovl \zeta}_2^2
+f_2f_3\zeta_2^2{\ovl \zeta}_1{\ovl \zeta}_2)
\label{cal-G}
\\
\phantom{{\cal G}(\xi ,\eta)x}
-\frac{1}{6}(f_2^2\zeta_1\zeta_2{\ovl \zeta}_1{\ovl \zeta}_2
+f_3^2\zeta_1\zeta_2{\ovl \zeta}_1{\ovl \zeta}_2)
-\frac{5}{48}(f_2^2\zeta_1^2{\ovl \zeta}_1^2
+f_3^2\zeta_2^2{\ovl \zeta}_2^2)
\nonumber
\\
\phantom{{\cal G}(\xi ,\eta)x}
-\frac{1}{8}(f_2^2\zeta_1^2{\ovl \zeta}_2^2
+f_2^2\zeta_2^2{\ovl \zeta}_1^2
+f_3^2\zeta_1^2{\ovl \zeta}_2^2
+f_3^2\zeta_2^2{\ovl \zeta}_1^2)
\nonumber
\\
\phantom{{\cal G}(\xi ,\eta)x}
+\frac{1}{16}(f_1f_3\zeta_1^2{\ovl \zeta}_2^2+f_1f_3\zeta_2^2{\ovl \zeta}_1^2
+f_2f_4\zeta_1^2{\ovl \zeta}_2^2+f_2f_4\zeta_2^2{\ovl \zeta}_1^2)
.
\nonumber
\end{eqnarray}
Note that if we choose $(f_h)$ to be
\begin{equation}
\label{f-HH}
f_1=f_3=0, \quad f_2=1, \quad f_4=\mu,
\end{equation}
${\cal G}$ becomes $G_{\mu}$, the BG-normal form
for the one-parameter H{\' e}non-Heiles Hamiltonian.
\par
We solve the degree-$4$ inverse problem
for the BG-normal form ${\cal G}$ in turn:
By ANFER, we have the following polynomial of degree-$4$
as the solution;
\begin{equation}
\label{cal-H}
{\cal H}(q,p)
=
\frac{1}{2}\sum_{j=1}^{2}
\left( p_j^2 + q_j^2 \right)
+
{\cal H}_3(q,p)
+
{\cal H}_4(q,p)
,
\end{equation}
with
\begin{equation}
\label{cal-H3}
\eqalign{
{\cal H}_3(q,p)=
a_1z_{1}^3
+a_2z_{1}^2z_{2}+a_3z_{1}z_{2}^2
+a_4z_{2}^3+a_5z_{1}^2{\ovl z}_{1}
+a_6z_{1}^2{\ovl z}_{2}
\\
\phantom{{\cal H}_3(q,p)=}
+a_7z_{1}z_{2}{\ovl z}_{1}+a_8z_{1}z_{2}{\ovl z}_{2}
+a_9z_{2}^2{\ovl z}_{1}+a_{10}z_{2}^2{\ovl z}_{2}
\\
\phantom{{\cal H}_3(q,p)=}
+{\ovl a}_1{\ovl z}_{1}^3
+{\ovl a}_2{\ovl z}_{1}^2{\ovl z}_{2}+{\ovl a}_3{\ovl z}_{1}{\ovl z}_{2}^2
+{\ovl a}_4{\ovl z}_{2}^3+{\ovl a}_5z_{1}{\ovl z}_{1}^2
+{\ovl a}_6z_{2}{\ovl z}_{1}^2
\\
\phantom{{\cal H}_3(q,p)=}
+{\ovl a}_7z_{1}{\ovl z}_{1}{\ovl z}_{2}
+{\ovl a}_8z_{2}{\ovl z}_{1}
{\ovl z}_{2}+{\ovl a}_9z_{1}{\ovl z}_{2}^2
+{\ovl a}_{10}z_{2}{\ovl z}_{2}^2 ,
}
\end{equation}
and
\begin{eqnarray}
\begin{array}{l}
\displaystyle{
{\cal H}_4(q,p)
}
\\
\displaystyle{
=
c_1z_{1}^4+c_2z_{1}^3z_{2}+c_3z_{1}^2z_{2}^2+c_4z_{1}z_{2}^3
}
\end{array}
\nonumber
\\
\phantom{=}
+c_5z_{2}^4+c_6z_{1}^3{\ovl z}_{1}
+c_7z_{1}^3{\ovl z}_{2}
+c_8z_{1}^2z_{2}{\ovl z}_{1}
\nonumber
\\
\phantom{=}
+c_9z_{1}^2z_{2}{\ovl z}_{2}
+c_{10}z_{1}z_{2}^2{\ovl z}_{1}
+c_{11}z_{1}z_{2}^2{\ovl z}_{2}
+c_{12}z_{2}^3{\ovl z}_{1}
+c_{13}z_{2}^3{\ovl z}_{2}
\nonumber
\\
\phantom{=}
+{\ovl c}_1{\ovl z}_{1}^4
+{\ovl c}_2{\ovl z}_{1}^3{\ovl z}_{2}
+{\ovl c}_3{\ovl z}_{1}^2{\ovl z}_{2}^2
+{\ovl c}_4{\ovl z}_{1}{\ovl z}_{2}^3
\nonumber
\\
\phantom{=}
+{\ovl c}_5{\ovl z}_{2}^4
+{\ovl c}_6z_{1}{\ovl z}_{1}^3
+{\ovl c}_7z_{2}{\ovl z}_{1}^3
+{\ovl c}_8z_{1}{\ovl z}_{1}^2{\ovl z}_{2}
\nonumber
\\
\phantom{=}
+{\ovl c}_9z_{2}{\ovl z}_{1}^2{\ovl z}_{2}
+{\ovl c}_{10}z_{1}{\ovl z}_{1}{\ovl z}_{2}^2
+{\ovl c}_{11}z_{2}{\ovl z}_{1}{\ovl z}_{2}^2
+{\ovl c}_{12}z_{1}{\ovl z}_{2}^3
+{\ovl c}_{13}z_{2}{\ovl z}_{2}^3
\nonumber
\\
\phantom{=}
+8(
a_6{\ovl a}_6z_{1}z_{2}{\ovl z}_{1}{\ovl z}_{2}
+a_9{\ovl a}_9z_{1}z_{2}{\ovl z}_{1}{\ovl z}_{2}
+a_5{\ovl a}_6z_{1}z_{2}{\ovl z}_{1}^2
\nonumber
\\
\phantom{====}
+a_6{\ovl a}_5z_{1}^2{\ovl z}_{1}{\ovl z}_{2}
+a_9{\ovl a}_{10}z_{2}^2{\ovl z}_{1}{\ovl z}_{2}
+a_{10}{\ovl a}_9z_{1}z_{2}{\ovl z}_{2}^2
)
\nonumber
\\
\phantom{=}
+6(
a_1{\ovl a}_1z_{1}^2{\ovl z}_{1}^2
+a_4{\ovl a}_4z_{2}^2{\ovl z}_{2}^2
+a_5{\ovl a}_5z_{1}^2{\ovl z}_{1}^2
+a_{10}{\ovl a}_{10}z_{2}^2{\ovl z}_{2}^2
)
\nonumber
\\
\phantom{=}
+4(
a_1{\ovl a}_2z_{1}^2{\ovl z}_{1}{\ovl z}_{2}
+a_8{\ovl a}_9z_{1}^2{\ovl z}_{2}^2
+a_3{\ovl a}_4z_{1}z_{2}{\ovl z}_{2}^2
+a_5{\ovl a}_8z_{1}z_{2}{\ovl z}_{1}{\ovl z}_{2}
\nonumber
\\
\phantom{====}
+a_6{\ovl a}_7z_{1}^2{\ovl z}_{2}^2
+a_6{\ovl a}_8z_{1}z_{2}{\ovl z}_{2}^2
+a_7{\ovl a}_9z_{1}^2{\ovl z}_{1}{\ovl z}_{2}
+a_7{\ovl a}_{10}z_{1}z_{2}{\ovl z}_{1}{\ovl z}_{2}
)
\nonumber
\\
\phantom{=}
+4(
a_2{\ovl a}_1z_{1}z_{2}{\ovl z}_{1}^2
+a_4{\ovl a}_3z_{2}^2{\ovl z}_{1}{\ovl z}_{2}
+a_8{\ovl a}_5z_{1}z_{2}{\ovl z}_{1}{\ovl z}_{2}
+a_7{\ovl a}_6z_{2}^2{\ovl z}_{1}^2
\nonumber
\\
\phantom{====}
+a_8{\ovl a}_6z_{2}^2{\ovl z}_{1}{\ovl z}_{2}
+a_9{\ovl a}_7z_{1}z_{2}{\ovl z}_{1}^2
+a_{10}{\ovl a}_7z_{1}z_{2}{\ovl z}_{1}{\ovl z}_{2}
+a_9{\ovl a}_8z_{2}^2{\ovl z}_{1}^2
)
\nonumber
\\
\phantom{=}
+\frac{8}{3}(
a_2{\ovl a}_2z_{1}z_{2}{\ovl z}_{1}{\ovl z}_{2}
+a_3{\ovl a}_3z_{1}z_{2}{\ovl z}_{1}{\ovl z}_{2}
)
\nonumber
\\
\phantom{=}
+2(
-a_6{\ovl a}_6z_{1}^2{\ovl z}_{1}^2
+a_7{\ovl a}_7z_{1}^2{\ovl z}_{1}^2
+a_8{\ovl a}_8z_{2}^2{\ovl z}_{2}^2
-a_9{\ovl a}_9z_{2}^2{\ovl z}_{2}^2
)
\label{cal-H4}
\\
\phantom{=}
+2(
a_1{\ovl a}_3z_{1}^2{\ovl z}_{2}^2
+a_2{\ovl a}_4z_{1}^2{\ovl z}_{2}^2
+a_5{\ovl a}_7z_{1}^2{\ovl z}_{1}{\ovl z}_{2}
-a_5{\ovl a}_9z_{1}^2{\ovl z}_{2}^2
\nonumber
\\
\phantom{====}
-a_6{\ovl a}_8z_{1}^2{\ovl z}_{1}{\ovl z}_{2}
-a_6{\ovl a}_{10}z_{1}^2{\ovl z}_{2}^2
+a_7{\ovl a}_8z_{1}z_{2}{\ovl z}_{1}^2
\nonumber
\\
\phantom{====}
+a_7{\ovl a}_8z_{2}^2{\ovl z}_{1}{\ovl z}_{2}
-a_7{\ovl a}_9z_{1}z_{2}{\ovl z}_{2}^2
+a_8{\ovl a}_{10}z_{1}z_{2}{\ovl z}_{2}^2
)
\nonumber
\\
\phantom{=}
+2(
a_3{\ovl a}_1z_{2}^2{\ovl z}_{1}^2
+a_4{\ovl a}_2z_{2}^2{\ovl z}_{1}^2
+a_7{\ovl a}_5z_{1}z_{2}{\ovl z}_{1}^2
-a_9{\ovl a}_5z_{2}^2{\ovl z}_{1}^2
\nonumber
\\
\phantom{====}
-a_8{\ovl a}_6z_{1}z_{2}{\ovl z}_{1}^2
-a_{10}{\ovl a}_6z_{2}^2{\ovl z}_{1}^2
+a_8{\ovl a}_7z_{1}^2{\ovl z}_{1}{\ovl z}_{2}
\nonumber
\\
\phantom{====}
+a_8{\ovl a}_7z_{1}z_{2}{\ovl z}_{2}^2
-a_9{\ovl a}_7z_{2}^2{\ovl z}_{1}{\ovl z}_{2}
+a_{10}{\ovl a}_8z_{2}^2{\ovl z}_{1}{\ovl z}_{2}
)
\nonumber
\\
\phantom{=}
+\frac{4}{3}(
a_2{\ovl a}_3z_{1}^2{\ovl z}_{1}{\ovl z}_{2}
+a_2{\ovl a}_3z_{1}z_{2}{\ovl z}_{2}^2
+a_3{\ovl a}_2z_{1}z_{2}{\ovl z}_{1}^2
+a_3{\ovl a}_2z_{2}^2{\ovl z}_{1}{\ovl z}_{2}
)
\nonumber
\\
\phantom{=}
+\frac{2}{3}(
a_2{\ovl a}_2z_{1}^2{\ovl z}_{1}^2
+a_3{\ovl a}_3z_{2}^2{\ovl z}_{2}^2
)
\nonumber
\\
\phantom{=}
-\frac{15}{16}(
f_1^2z_{1}^2{\ovl z}_{1}^2
+f_4^2z_{2}^2{\ovl z}_{2}^2
)
-\frac{3}{4}(
f_1f_3z_{1}z_{2}{\ovl z}_{1}{\ovl z}_{2}
+f_2f_4z_{1}z_{2}{\ovl z}_{1}{\ovl z}_{2}
)
\nonumber
\\
\phantom{=}
-\frac{5}{8}(
f_1f_2z_{1}^2{\ovl z}_{1}{\ovl z}_{2}
+f_1f_2z_{1}z_{2}{\ovl z}_{1}^2
+f_3f_4z_{1}z_{2}{\ovl z}_{2}^2
+f_3f_4z_{2}^2{\ovl z}_{1}{\ovl z}_{2}
)
\nonumber
\\
\phantom{=}
-\frac{5f_2f_3}{24}(
z_{1}^2{\ovl z}_{1}{\ovl z}_{2}
+z_{1}z_{2}{\ovl z}_{1}^2
+z_{1}z_{2}{\ovl z}_{2}^2
+z_{2}^2{\ovl z}_{1}{\ovl z}_{2}
)
\nonumber
\\
\phantom{=}
-\frac{1}{6}(
f_2^2z_{1}z_{2}{\ovl z}_{1}{\ovl z}_{2}
f_3^2z_{1}z_{2}{\ovl z}_{1}{\ovl z}_{2}
)
\nonumber
\\
\phantom{=}
-\frac{1}{8}(
f_2^2z_{1}^2{\ovl z}_{2}^2
+f_2^2z_{2}^2{\ovl z}_{1}^2
+f_3^2z_{1}^2{\ovl z}_{2}^2
+f_3^2z_{2}^2{\ovl z}_{1}^2
)
-\frac{5}{48}(
f_2^2z_{1}^2{\ovl z}_{1}^2
+f_3^2z_{2}^2{\ovl z}_{2}^2
)
\nonumber
\\
\phantom{=}
+\frac{1}{16}(
f_1f_3z_{2}^2{\ovl z}_{1}^2
+f_1f_3z_{1}^2{\ovl z}_{2}^2
+f_2f_4z_{1}^2{\ovl z}_{2}^2
+f_2f_4z_{2}^2{\ovl z}_{1}^2
),
\nonumber
\end{eqnarray}
where $a_h$ ($h=1,\cdots , 10$) and $c_{\ell}$
($\ell = 1, \cdots , 13$) are the complex-valued
parameters chosen arbitrarily,
and $f_k$ ($k=1,\cdots,4$) the real-valued
parameters in ${\cal P}^{(3)}(q)$ (see (\ref{P3})).
Namely, we have 46-degree-of freedom in the solution, ${\cal H}$,
of the inverse problem of the PHOCP with $(f_k)$ fixed.
Note that if $(f_k)$ are chosen to be (\ref{f-HH})
then ${\cal H}$ becomes $H_{\mu}$ given by (\ref{H-HH})-(\ref{H4-HH}).
Further, if $(a_h)$, $(c_{\ell})$ and $(f_k)$ are chosen to be
(\ref{special1-HH}), (\ref{special2-HH}) and (\ref{f-HH}),
respectively, then ${\cal H}$ becomes $K_{\mu}$ given by
(\ref{K-HH}).
After (\ref{cal-H})-(\ref{cal-H4}),
one might understand more than after (\ref{H-HH})-(\ref{H4-HH})
the necessity of computer algebra in the inverse problem.
\subsection{The Bertrand-Darboux integrability condition}
We wish to find the condition for $(a_h)$, $(c_{\ell})$ and $(f_k)$
to bring ${\cal H}$ into the PHOQP-Hamiltonian
${\cal F}^{(4)}(q,p)$ defined by (\ref{F3-F4}) with (\ref{P4}).
In order to make ${\cal H}_3(q,p)$ vanish (see (\ref{cal-H3})),
we have to choose $(a_h)$ to be (\ref{vanish-H3}).
Our next task is then to bring ${\cal H}_4\vert_{a=0}(q,p)$
into a certain ${\cal P}^{(4)}(q)$.
Let $\nu_n$ ($n=1, \cdots ,5$) be real-valued
parameters. Then we see that ${\cal H}_4\vert_{a=0}$ takes the form
${\cal P}^{(4)}(q)$ if and only if the following identities of $z$
hold true for a non-vanishing $((c_h), (\nu_n))$;
\begin{eqnarray}
\nu_1 q_1^4
=
c_1 z_1^4 +c_6 z_1^3{\ovl z}_1
-\frac{45f_1^2+5f_2^2}{48}z_1^2{\ovl z}_1^2
+{\ovl c}_6z_1{\ovl z}_1^3 + {\ovl c}_1{\ovl z}_1^4
,
\nonumber \\
\nu_2 q_2^4
=
c_5 z_2^4 +c_{13} z_2^3{\ovl z}_2
-\frac{45f_4^2+5f_3^2}{48}z_2^2{\ovl z}_2^2
+{\ovl c}_{13}z_2{\ovl z}_2^3 + {\ovl c}_5{\ovl z}_1^4
,
\nonumber \\
\nu_3 q_1^3 q_2
=
c_2z_1^3z_2+c_7z_1^3{\ovl z}_2+c_8z_1^2z_2{\ovl z}_1
\nonumber \\
\phantom{xxxxxxx}
+{\ovl c}_2{\ovl z}_1^3{\ovl z}_2
+{\ovl c}_7{\ovl z}_1^3z_2
+{\ovl c}_8{\ovl z}_1^2{\ovl z}_2z_1
\nonumber \\
\phantom{xxxxxxx}
-\frac{5}{24}
(3f_1f_2+f_2f_3)
(z_1^2{\ovl z}_1{\ovl z}_2 +z_1z_2{\ovl z}_1^2)
,
\nonumber \\
\nu_4 q_1 q_2^3
=
c_4z_1z_2^3+c_{11}z_1z_2^2{\ovl z}_2
+c_{12}z_2^3{\ovl z}_1
\label{identity-PHOCP}
\\
\phantom{xxxxxxx}
+{\ovl c}_{12}z_1{\ovl z}_2^3
+{\ovl c}_{11}z_2{\ovl z}_1{\ovl z}_2^2
+{\ovl c}_4{\ovl z}_1{\ovl z}_2^3
\nonumber \\
\phantom{xxxxxxx}
-\frac{5}{24}
(3f_3f_4+f_2f_3)
(z_2^2{\ovl z}_1{\ovl z}_2 +z_1z_2{\ovl z}_2^2)
,
\nonumber \\
\nu_5 q_1^2q_2^2
=
c_3 z_1^2 z_2^2 + c_9 z_1^2 z_2 {\ovl z}_2
+c_{10}z_1z_2^2{\ovl z}_1
\nonumber \\
\phantom{xxxxxxx}
+{\ovl c}_{10}z_1{\ovl z}_1{\ovl z}_2^2
+{\ovl c}_9 z_2 {\ovl z}_1^2 {\ovl z}_2
+{\ovl c}_3 {\ovl z}_1^2 {\ovl z}_2^2
\nonumber \\
\phantom{xxxxxxx}
+\frac{1}{16}(f_1f_3+f_2f_4-2f_2^2-2f_3^2)
(z_1^2{\ovl z}_2^2 +{\ovl z}_1^2z_2^2)
\nonumber \\
\phantom{xxxxxxx}
-\frac{1}{12}
(9f_1f_3+9f_2f_4+2f_2^2+2f_3^2)
z_1z_2{\ovl z}_1{\ovl z}_2 ,
\nonumber
\end{eqnarray}
where $\nu_n$ ($n=1, \cdots , 5$) are real-valued parameters
and $q_j=(z_j + {\ovl z}_j )/2$ ($j=1,2$).
As a necessary and sufficient condition for
the identities to hold true from the first to the fourth, we have
\begin{eqnarray}
\nu_1=16c_1 = 4c_6 = -\frac{5}{18}(9f_1^2+f_2^2),
\nonumber \\
\nu_2=16c_5 = 4c_{13} = -\frac{5}{18}(9f_4^2+f_3^2),
\nonumber \\
\nu_3=16c_2=16c_7=\frac{16}{3}c_8
=
-\frac{10}{9}f_2(3f_1+f_3)
,
\label{c-nu-1}
\\
\nu_4=16c_4=16c_{12}=\frac{16}{3}c_{11}
=
-\frac{10}{9}f_3(3f_4+f_2).
\nonumber
\end{eqnarray}
As for the fifth identity, we see that it holds true if and only if
the following overdetermined system of equations,
\begin{equation}
\label{eq-c-nu}
\eqalign{
\nu_5=16c_3=8c_9=8c_{10},
\\ 
\nu_5
=
(f_1f_3+f_2f_4)-2(f_2^2+f_3^2)
,
\\
\nu_5
=
-3(f_1f_3+f_2f_4)-\frac{2}{3}(f_2^2+f_3^2),
}
\end{equation}
admits a solution.
Surprisingly, the Bertrand-Darboux integrability
condition (\ref{BDIC-P3-a}) for PHOCP's comes out as a necessary and
sufficient condition for (\ref{eq-c-nu}) to admit a solution~!
Under (\ref{BDIC-P3-a}), equation (\ref{eq-c-nu}) admits the solution,
\begin{equation}
\label{c-nu-2}
\nu_5=16c_3=8c_9=8c_{10}
=
(f_1f_3+f_2f_4)-2(f_2^2+f_3^2),
\end{equation}
which is combined with (\ref{vanish-H3}) and (\ref{c-nu-1})
to bring ${\cal H}$ into the PHOQP-Hamiltonian,
\begin{equation}
\label{cal-Q}
\eqalign{
{\cal Q}=\frac{1}{2}\sum_{j=1}^2 (p_j^2 + q_j^2)
\\
\phantom{{\cal Q}=}
-\frac{5}{18}(9f_1^2+f_2^2)q_1^4
-\frac{10}{9}(3f_1+f_3)f_2 q_1^3q_2
-\frac{5}{3}(f_2^2+f_3^2)q_1^2q_2^2
\\
\phantom{{\cal Q}=}
-\frac{10}{9}(3f_4+f_2)f_3 q_1q_2^3
-\frac{5}{18}(9f_4^2+f_3^2)q_2^4
,
}
\end{equation}
where $(f_h)$ are subject to the BDIC (\ref{BDIC-P3-a}).
We note here that the coefficient of $q_1^2q_2^2$ in (\ref{cal-Q})
is obtained by combining (\ref{c-nu-2}) with (\ref{BDIC-P3-a}).
To summarize, we have the following.
\begin{theorem}
\label{PHOQP-general}
The perturbed harmonic-oscillator Hamiltonian ${\cal F}^{(3)}$ with
a homogeneous cubic polynomial potential shares its BG-normal
form with 
the perturbed harmonic-oscillator Hamiltonian ${\cal F}^{(4)}$
with a homogeneous-quartic potential up to degree-$4$
if and only if the PHOCP-Hamiltonian ${\cal F}^{(3)}$ satisfies
the Bertrand-Darboux integrability condition (\ref{BDIC-P3-a}).
Under (\ref{BDIC-P3-a}), the PHOQP-Hamiltonian ${\cal F}^{(4)}$
sharing its BG-normal form with that PHOCP-Hamiltonian ${\cal F}^{(3)}$
is equal to ${\cal Q}$ given by (\ref{cal-Q}). 
\end{theorem}
We are now in a position to show the integrability of the PHOQP
with ${\cal Q}$ subject to (\ref{BDIC-P3-a}).
It is easy to see that ${\cal Q}$ is given by (\ref{F3-F4}) with (\ref{P4})
under the substitution,
\begin{eqnarray}
g_1= -\frac{5}{18}(9f_1^2+f_2^2),
\quad
g_2=-\frac{10}{9}(3f_1+f_3)f_2 ,
\quad
g_3=-\frac{5}{3}(f_2^2+f_3^2) ,
\nonumber \\
g_4=-\frac{10}{9}(3f_4+f_2)f_3 , 
\quad
g_5=-\frac{5}{18}(9f_4^2+f_3^2) .
\label{g-f}
\end{eqnarray}
A long but straightforward calculation shows that $(g_{\ell})$ given
by (\ref{g-f}) with (\ref{BDIC-P3-a}) satisfy the BDIC (\ref{BDIC-P4-b}),
so that the PHOQP with ${\cal Q}$ is integrable.
\begin{theorem}
\label{theorem-BDIC-CP-QP}
If the perturbed harmonic oscillators with a homogeneous-cubic polynomial
potential and with a homogeneous-quartic
polynomial potential share the same BG-normal form up to degree-$4$,
then both oscillators are integrable in the sense that they satisfy
the Bertrand-Darboux integrability condition.
\end{theorem}
\section{Concluding Remarks}
As is shown in the previous section,
the new deep relation between the BDIC for PHOCP's and that for PHOQP's
are found (see theorems~\ref{PHOQP-general} and \ref{theorem-BDIC-CP-QP}).
The results obtained in the present paper are expected to provide several
interesting subjects, which are listed below.
\begin{description}
\item[{(1)}]
A further generalization of theorems~\ref{PHOQP-general} and \ref{theorem-BDIC-CP-QP}
will be worth studying: A conjecture is posed
as follows, which is investigated now.
\begin{conjecture}
Let ${\cal F}^{(r)}(q,p)$ and ${\cal F}^{(2r-2)}(q,p)$
($r=3,4,\cdots$)
denote the perturbed harmonic oscillator Hamiltonians of the form
\begin{eqnarray}
\label{Fr}
{\cal F}^{(r)}(q,p)
=
\frac{1}{2}\sum_{j=1}^{2}(p_j^2+q_j^2)
+
V^{(r)}(q),
\\
\label{F2r-2)}
{\cal F}^{(2r-2)}(q,p)
=
\frac{1}{2}\sum_{j=1}^{2}(p_j^2+q_j^2)
+
V^{(2r-2)}(q),
\end{eqnarray}
where $V^{(2r-2)}$ is a homogeneous-polynomial potential of degree-$(2r-2)$
and $V^{(r)}$ of degree-$r$ subject to
\begin{equation}
\label{Vr}
V^{(r)}(q) \in \image D_{q,\eta}^{(r)}.
\end{equation}
If the perturbed harmonic oscillators with the
Hamiltonian ${\cal F}^{(r)}$ and with ${\cal F}^{(2r-2)}$
share a BGNF up to degree-$(2r-2)$ then both oscillators are integrable
in the sense that they satisfy the Bertrand-Darboux integrability
condition.
\end{conjecture}
\item[{(2)}]
As is pointed out after theorem~\ref{PHOQP-HH},
the Hamiltonians, $K_{1/3}$ and $Q$ have a significant feature
other than integrability; they are separable in $q_1 \pm q_2$.
Such a separability can be found also in
${\cal F}^{(3)}$ with (\ref{BDIC-P3-a}) and
${\cal Q}$, the generalization of $K_{1/3}$ and $Q$, respectively.
In fact, we can find, from theorem~\ref{PHOQP-general}, the PHOCP's with
\begin{equation}
\label{P3-special}
{\cal P}^{(3)}(q)=a(q_1+q_2)^3 + b (q_1 -q_2)^3
\quad (a,b \in {\bf R}),
\end{equation}
as special cases of PHOCP's subject to the  BDIC (\ref{BDIC-P3-a}),
which cover all the PHOCP-Hamiltonians separable in $q_1 \pm q_2$
(see Perelomov 1990).
Further, theorem~\ref{PHOQP-general} implies that each of the separable
PHOCP's (subject to (\ref{P3-special})) shares the same BG-normal form with
the PHOQP with 
\begin{equation}
\label{P4-special}
{\cal P}^{(4)}(q)=-5 \{ (a^2 (q_1+q_2)^4 + b^2 (q_1 -q_2)^4 \},
\end{equation}
which is also known to be separable in $q_1 \pm q_2$ (Perelomov 1990).
It is worth noting that all the PHOQP's with (\ref{P4-special}) cover
a quarter of the class of all the PHOQP-Hamiltonians separable in
$q_1 \pm q_2$ in view of $a^2 , b^2 \geq 0$.
Thus we reach through theorem~\ref{PHOQP-general} to the four-to-one
correspondence between the PHOCP's with (\ref{P3-special}) and PHOQP's
with (\ref{P4-special}) separable in $q_1 \pm q_2$.
Since ${\cal F}^{(3)}$ with (\ref{BDIC-P3-a}) and ${\cal Q}$
are thought to include several classes of Hamiltonians separable in
several coordinate systems other than $q_1 \pm q_2$,
the separability will be worth studying extensively
from the BG-normalization viewpoint in future.
\item[{(3)}]
The perturbed oscillators referred to in theorems~\ref{PHOQP-general} and
\ref{theorem-BDIC-CP-QP} are expected to provide good examples
of the quantum bifurcation in the BG-normalized Hamiltonian systems
(Uwano 1994, 1995, 1998, 1999):
Since the perturbed oscillators referred to in
theorems~\ref{PHOQP-general} and \ref{theorem-BDIC-CP-QP}
are integrable, their quantum spectra are expected to be obtained exactly.
Then we will be able to think of whether or not the quantum bifurcation
in the BG-normalized Hamiltonian system for those oscillators
approximates the bifurcation in these oscillators in a good extent.
\item[{(4)}]
In section~1, two approaches to the BDIC (or BD-theorem) have been
mentioned.
As another approach to the BDIC,
the work of Yamaguchi and Nambu (1998) is worth pointing out,
in which the BDIC for the PHOCP's and the PHOQP's (see
Lemma~\ref{lemma-YN}) came out from the renormalization of
Hamiltonian equations.
It will be
an interesting problem in future to study a relation between
the BG-normalization and the renormalization.
\item[{(5)}]
As is easily seen,  the solution (\ref{cal-H}, \ref{cal-H3}, \ref{cal-H4})
of the inverse problem for ${\cal G}$ admits fifty real-valued
parameters. We may hence expect to obtain
other integrable systems, so-called the electromagnetic
type (Hietarinta 1987), for example.
\end{description}
On closing this section, we wish to mention
of the role of computer algebra in the present paper:
Without computer algebra, it would have been very difficult to find
theorems~\ref{PHOQP-HH}, \ref{theorem-BDIC-HH}, \ref{PHOQP-general}
and \ref{theorem-BDIC-CP-QP}.
\par
\ack
The author wishes to thank Prof.~T.~Iwai and Dr.~Y.~Y.~Yamaguchi
at Kyoto University, and Dr.~S.~I.~Vinitsky at the Joint
Institute for Nuclear Research, Russia, for their valuable comments.
Thanks are also to the referees for their valuable comments to
the earlier manuscript of this paper. 
The present work is partly supported by the Grant-in-Aid Exploratory
Research no.~11875022 from the Ministry of Education, Science and
Culture, Japan. 
\section*{References}
\begin{harvard}
\item[]
Arnold V I 1980
{\it Mathematical Methods of Classical Mechanics} 
(New York: Springer-Verlag) p~215
\item[]
Chekanov~N~A Hongo~M Rostovtsev~V~A Uwano~Y and Vinitsky~S~I
1998 {\it Physics of Atomic Nuclei} {\bf 61} 2029
\item[]
Chekanov~N~A Rostovtsev~V~A Uwano~Y and Vinitsky~S~I
2000 {\it Comp.~Phys.~Comm.} {\bf 126} 47
\item[]
Cushman~R 1982 {\it Proc.~R.~Soc.} {\bf A382} 361
\item[]
Darboux~G 1901 {\it Archives Neerlandaises} (ii) {\bf 6} 371
\item[]
Goldstein H 1950 {\it Classical Mechanics} 2nd~ed.
(Reading: Addison-Wesley) p~241  
\item[]
Grosche~C Pogosyan~G~S and Sissakian~A~N 1995
{\it Fortschr. Phys.} {\bf 43} 453
\item[]
Gustavson~F~G 1966 {\it Astronomical Journal} {\bf 71} 670
\item[]
Hietarinta~J 1987 {\it Phys. Rep.} {\bf 147} 87
\item[]
Kummer~M 1976 {\it Comm. Math. Phys.} {\bf 48} 53
\item[]
Marshall~I and Wojciechowski~S 1988 {\it J.~Math.~Phys.} {\bf 29}
1338
\item[]
Moser~J~K 1968 {\it Lectures on Hamiltonian Systems} Memoirs of
A.M.S. {\bf 81} (Providence: A.M.S.) p~10
\item[]
Perelomov~A~M 1990 {\it Integrable Systems of Classical Mechanics
and lie Algebras} vol.1 (Basel: Birkh{\" a}user-Verlag)
p~81
\item[]
Spivak~M 1965 {\it Calculus on Manifolds} (New York: Benjamin)
p~41 
\item[]
Uwano~Y 1994 {\it J. Phys. Soc. Jpn} {\bf 63 Suppl.~A} 31
\item[]
\dash 1995 {\it J. Phys.} {\bf A28} 2041
\item[]
\dash 1998 {\it Int. J. of Bifurcation and Chaos} {\bf 8} 941
\item[]
\dash 1999 {\it Rep. Math. Phys.} {\bf 44} 267
\item
\dash 2000 http://yang.amp.i.kyoto-u.ac.jp/\~{}uwano/
\item[]
Uwano~Y Chekanov~N Rostovtsev~V and Vinitsky~S 1999
{\it Computer Algebra in Scientific Computing}
ed Ganzha~V~G et al (Berlin: Springer-Verlag) 441
\item[]
Whittaker~E~T 1937 {\it A Treatise on the Analytical Dynamics
of Particles and Rigid Bodies} 4th ed. (Cambridge: Cambridge U.P.)
p~332
\item[]
Yamaguchi~Y~Y and Nambu~Y 1998 {\it Prog. Theor. Phys.} {\bf 100}
199
\end{harvard}
\appendix
\section{Mathematical basis for coding ANFER}
In this appendix, we describe in mathematical terminology
the algorithm of ANFER, a symbolic computing programme,
for the degree-$4$ inverse problem on REDUCE~3.6 coded by the author.
Although only a primitive prototype exists,
those who are interested in ANFER will be able to see its source-code
at Uwano (2000), http://yang.amp.i.kyoto-u.ac.jp/\~{}uwano/.
After the algorithm, we present a key lemma supporting
the algorithm.
The general-degree case has been reported
in Uwano et al (1999) briefly, and will be discussed in detail
in a future paper by the author.
\par
\subsection{Setting-up}
The setting-up of ANFER is made as follows for the degree-$2\delta$ inverse
problem.
Let ${\cal S}^{(h)}(\xi^{(h)}, \eta^{(h-1)})$
($h=3, \cdots , 2\delta$) be the third-type generating functions of
the form
\begin{equation}
\label{def-calS}
{\cal S}^{(h)}(\xi^{(h)},\eta^{(h-1)})
=
-\sum_{j=1}^{2}\xi^{(h)}\eta^{(h-1)}
-{\cal S}_{h}(\xi^{(h)},\eta^{(h-1)})
\quad
(h=3, \cdots ,2\delta),
\end{equation}
where each ${\cal S}_{h}$ are the homogeneous polynomial
of degree-$h$. With ${\cal S}^{(h)}$, we associate the canonical
transformations,
\begin{equation}
\label{tau-h}
\tau_{h}: (\xi^{(h-1)},\eta^{(h-1)})
           \rightarrow ( \xi^{(h)},\eta^{(h)})
\quad
(h=3, \cdots , 2\delta),
\end{equation}
with
\begin{equation}
\label{transf-h}
\xi^{(h-1)}
=
-\frac{\partial {\cal S}^{(h)}}{\partial \eta^{(h-1)}},
\quad
\eta^{(h)}
=
-\frac{\partial {\cal S}^{(h)}}{\partial \xi^{(h)}}
\quad (h=3, \cdots , 2\delta ).
\end{equation}
From a given BG-normal form $G$ of the form (\ref{G}),
we define the initial Hamiltonian $H^{(2)}$
to be the degree-$2\delta$ polynomial form,
\begin{equation}
\label{H2}
H^{(2)}(\xi^{(2)},\eta^{(2)})
=
\frac{1}{2}\sum_{j=1}^{2}\left(
(\eta_j^{(2)})^2 + (\xi_j^{(2)})^2 
\right)
+ \sum_{k=3}^{2\delta}G_{k}(\xi^{(2)},\eta^{(2)}).
\end{equation}
In ANFER, the degree-$2\delta$ inverse problem,
\begin{eqnarray}
\label{defeq-iterate}
H^{(h)}(\xi^{(h)},
-\frac{\partial {\cal S}^{(h)}}{\partial \xi^{(h)}}
)
=
H^{(h-1)}(
-\frac{\partial {\cal S}^{(h)}}{\partial \eta^{(h-1)}},
\eta^{(h-1)}
),
\\
\label{cond-calS}
{\cal S}^{(h)}(\xi^{(h)},\eta^{(h)}) 
\in
\image D_{\xi^{(h)},\eta^{(h-1)}}
\quad \mbox{with (\ref{def-calS})}
,
\end{eqnarray}
for $H^{(h-1)}$ are solved recursively with $h=3, \cdots , 2\delta$.
The resultant $H^{(2\delta)}$ provides the solution $H$ of the
inverse problem for the BG-normal form $G$ (Uwano et al 1999).
\par
\subsection{Algorithm for the degree-$4$ inverse problem}
We show how (\ref{defeq-iterate}, \ref{cond-calS}) is solved
in the degree-$4$ case:
\par
\smallskip\noindent
{\bf Step-I}: Solving (\ref{defeq-iterate}, \ref{cond-calS})
with $h=3$
\newline
Equating the homogeneous-parts of degree-$2$, $3$ and $4$ in
(\ref{defeq-iterate}) with $h=3$, we have
\begin{eqnarray}
\label{H3-2}
H_2^{(3)}(\xi^{(3)},\eta^{(2)})
=
H_2^{(2)}(\xi^{(3)},\eta^{(2)}),
\\ \noalign{\vskip 6pt}
\label{H3-3}
H_3^{(3)}(\xi^{(3)},\eta^{(2)})
-(D_{\xi^{(3)},\eta^{(2)}}{\cal S}_{3})(\xi^{(3)},\eta^{(2)})
=H_3^{(2)}(\xi^{(3)},\eta^{(2)})
\\ \noalign{\vskip 9pt}
\nonumber
H_4^{(3)}(\xi^{(3)},\eta^{(2)})
\\
\label{H3-4}
=
H_4^{(2)}(\xi^{(3)},\eta^{(2)})
-\sum_{j=1}^{2} \left (
\frac{1}{2}
\left(
\frac{\partial {\cal S}_3}{\partial \xi^{(3)}_j} 
\right)^2
+
\frac{\partial {\cal S}_3}{\partial \xi_j^{(3)}}
\left.
\frac{\partial H_3^{(3)}}{\partial \eta^{(3)}_j}
\right\vert_{(\xi^{(3)},\eta^{(2)})}
\right.
\\
\nonumber
\phantom{
H_4^{(3)}(\xi^{(3)},\eta^{(2)})
=\sum_{j=1}^{2}xxxxxx
}
\left.
-
\frac{1}{2}
\left(
\frac{\partial {\cal S}_3}{\partial \eta^{(2)}_j} 
\right)^2
-
\frac{\partial {\cal S}_3}{\partial \eta_j^{(2)}}
\left.
\frac{\partial H_3^{(2)}}{\partial \xi_j^{(2)}}
\right\vert_{(\xi^{(3)},\eta^{(2)})}
\right)
.
\end{eqnarray}
In a similar way to (\ref{eq-inv}-\ref{solution-inv-image-S}),
(\ref{H3-3}) is solved to be
\begin{equation}
\label{solution-H3-3}
\eqalign{
H_{3}^{(3)}\phantom{}^{\ker}(\xi^{(3)}, \eta^{(2)}) 
=
H_{3}^{(2)}\phantom{}^{\ker}(\xi^{(3)}, \eta^{(2)}) =0,
\\
(H_{3}^{(3)})^{\supim}(\xi^{(3)}, \eta^{(2)})
\in \image D_{\xi^{(3)},\eta^{(2)}}^{(3)}:
\; \mbox{chosen arbitrarily},
\\
{\cal S}_{3}(\xi^{(3)},\eta^{(2)})
\\
=
\left(
\left.
D_{q,\eta}^{(3)}
\right\vert_{\supim D_{q,\eta}^{(3)}}^{-1}
H_3^{(3)}\phantom{}^{\supim}\vert_{(q,\eta)})
\right)
(\xi^{(3)},\eta^{(2)}).
}
\end{equation}
The $H_4^{(3)}$ is given by (\ref{H3-4}) with the substitution
(\ref{solution-H3-3}) into $H_3^{(3)}$ and ${\cal S}_3$.
\par\smallskip
\noindent
{\bf Step-II}: Solving (\ref{defeq-iterate}, \ref{cond-calS})
with $h=4$
\newline 
Equating the homogeneous-parts of degree-$2$, $3$ and $4$
in (\ref{defeq-iterate}) with $h=4$, we have
\begin{eqnarray}
\label{H4-2,3}
H_{k}^{(4)}(\xi^{(4)},\eta^{(3)})
=
H_{k}^{(3)}(\xi^{(4)}, \eta^{(3)})
\quad
(k=2,3),
\\
\label{H4-4}
H_4^{(4)}(\xi^{(4)},\eta^{(3)})
-(D_{\xi^{(4)},\eta^{(3)}}{\cal S}_{4})(\xi^{(4)},\eta^{(3)})
=H_4^{(3)}(\xi^{(4)},\eta^{(3)}) .
\end{eqnarray}
In a similar way to (\ref{eq-inv}-\ref{solution-inv-image-S}),
(\ref{H4-4}) is solved to be
\begin{equation}
\label{solution-H4-4}
\eqalign{
H_{4}^{(4)}\phantom{}^{\ker}(\xi^{(4)}, \eta^{(3)}) 
= 
H_{3}^{(4)}\phantom{}^{\ker}(\xi^{(4)}, \eta^{(3)}) 
\\
H_{4}^{(4)}\phantom{}^{\supim}(\xi^{(4)}, \eta^{(3)})
\in \image D_{\xi^{(4)},\eta^{(3)}}^{(4)}:
\; \mbox{chosen arbitrarily},
\\
{\cal S}_{4}(\xi^{(4)},\eta^{(3)})
\\
=
\left(
\left. 
D_{q,\eta}^{(4)}
\right\vert_{\supim D_{q,\eta}^{(4)}}^{-1}
\left(
H_4^{(4)}\phantom{}^{\supim}\vert_{(q,\eta)}
 -H_4^{(3)}\phantom{}^{\supim}\vert_{(q,\eta)}
\right)
\right)
(\xi^{(4)},\eta^{(3)}).
}
\end{equation}
\par
We are now in a position to show that $H^{(4)}(q,p)$ and
$
{\cal S}(q,\eta)= -\sum_{j=1}^{2}q_j\eta_j -\sum_{k=3}^{4}{\cal S}_k (q,\eta)
$
are identical with $H(q,p)$ and $S(q,\eta)$, respectively,
the solution of the degree-$4$ inverse problem for $G$.
Equations (\ref{Psi}), $G_{3}(q,\eta)=0$ and $\Psi_{3}(q,\eta)=0$
(see the line above (\ref{Psi}))
are put together with (\ref{solution-inv-ker}, \ref{solution-inv-image-H},
\ref{solution-inv-image-S}) to show that
$H_{3}(q,p)$ is equal to $H_3^{(3)}(q,p)$,
so that we have $S_3(q,\eta)={\cal S}_3(q,\eta)$.
Under $H_3(q,p)=H_3^{(3)}(q,p)$ and $S_3(q,\eta)={\cal S}_3(q,\eta)$,
we see that the second term in the r.h.s. of (\ref{H3-4}) is equal to
$-\Psi_4$ given by (\ref{Psi}) with $(q,\eta)=(\xi^{(3)},\eta^{(2)})$.
This fact and (\ref{H4-4}) are put together with (\ref{solution-H4-4})
to show that $H_{4}^{(4)}(q,p)$ and ${\cal S}_{4}(q,\eta)$
are identical with $H_{4}(q,p)$ and $S_{4}(q,\eta)$,
respectively.
\par
What is characteristic of the algorithm above is that
the procedure of solving (\ref{eq-inv}) with $k=4$ is divided
into a pair of steps, Step-I and Step-II (extending to the degree-$2\delta$
case, we have $2\delta-2$ steps). Although it is not so significant
in the inverse problem of lower-degree (like degree-$4$),
such a division will contribute a lot to the reduction of the
memory-size required through computation.
\par
\subsection{Composition of canonical transformations}
We wish to give an explicit expression of the generating
function associated with the composition
$\tau_{4} \circ \tau_{3}$ of canonical transformations,
$\tau_{3}$ and $\tau_{4}$, which support mathematically
the algorithm given above. The general-degree version
can be found in Uwano et al (1999) without a proof.
Such an expression is really important because
little is known  explicitly of the composition of {\it non-infinitesimal}
canonical although well-known is that
of infinitesimal ones (Goldstein 1950).
\par\vskip 3pt\noindent
{\it Lemma A.1}\quad
The composition
$
\tau_{4} \circ \tau_3 :(\xi^{(2)},\eta^{(2)})
\rightarrow (\xi^{(4)},\eta^{(4)})
$
of the
canonical transformations $\tau_{3}$ and $\tau_{4}$ defined by (\ref{tau-h})
is associated with the third-type generating function
\begin{equation}
\label{tilde-calS}
{\tilde {\cal S}}(\xi^{(4)}, \eta^{(2)})
=
\sum_{j=1}^{2}
{\tilde \xi}_j {\tilde \eta}_j
+{\cal S}^{(3)}({\tilde \xi}, \eta^{(2)})
+{\cal S}^{(4)}(\xi^{(4)}, {\tilde \eta}),
\end{equation}
where ${\tilde \xi}$ and ${\tilde \eta}$ are
the functions of $\xi^{(4)}$ and $\eta^{(2)}$
uniquely determined to satisfy
\begin{equation}
\label{implicit-funct}
{\tilde \xi}=
-\frac{\partial {\cal S}^{(4)}}{\partial \eta^{(3)}}
(\xi^{(4)}, {\tilde \eta}),
\quad
{\tilde \eta}=
-\frac{\partial {\cal S}^{(3)}}{\partial \xi^{(3)}}
({\tilde \xi}, \eta^{(2)}),
\end{equation}
around $(\xi^{(4)}, \eta^{(2)})=0$.
The ${\tilde {\cal S}}$ admits the expansion,
\begin{eqnarray}
\label{expansion-tilde-calS}
{\tilde {\cal S}}(q,\eta)
=
-\sum_{j=1}^{2}q_j \eta_{j} -{\cal S}_{3}(q,\eta)
-{\cal S}_{4}(q,\eta) + o_{4}(q,\eta),
\\
\label{o4}
\frac{o_{4}(q,\eta)}
{\sqrt{\sum_{j=1}^{2}(q_j^2 +\eta_j^2)}^{4}}
 \rightarrow 0
\qquad (\sqrt{\sum_{j=1}^{2}(q_j^2 +\eta_j^2)}
         \rightarrow 0 ).
\end{eqnarray}
\smallskip
{\it Proof}:\quad
Since the third-type generating functions, ${\cal S}^{(h)}$'s,
satisfy
\begin{equation}
\label{def-eq-type-3-gen}
-\xi^{(h-1)}d\eta^{(h-1)}-\eta^{(h)}d\xi^{(h)}
=d{\cal S}^{(h)}
\quad
(h=3,4)
\end{equation}
 (Goldstein 1950), we have
\begin{equation}
\label{sum-defeq}
\eqalign{
-\xi^{(2)}d\eta^{(2)}-\eta^{(4)}d\xi^{(4)}
\\
=
d\left\{
(\sum_{j=1}^2 \xi_j^{(3)}\eta_j^{(3)}) 
+ {\cal S}^{(3)} (\xi^{(3)},\eta^{(2)})
+ {\cal S}^{(4)} (\xi^{(4)},\eta^{(3)})
\right\}.
}
\end{equation}
We express $\xi^{(3)}$ and $\eta^{(3)}$ in terms of
${\tilde \xi}$ and ${\tilde \eta}$ in turn:
Applying the implicit function theorem (Spivak 1965)
to the map,
\begin{equation}
\label{implicit-map}
\begin{array}{l}
\displaystyle{
\tau :(\xi^{(3)}, \eta^{(3)}, \eta^{(2)}, \xi^{(4)}) \in {\bf R}^4
}
\\ \noalign{\vskip 9pt}
\displaystyle{
\mapsto
\left(
\xi^{(3)}
+\frac{\partial {\cal S}^{(4)}}{\partial \eta^{(3)}}
( \xi^{(4)} , \eta^{(3)}), 
\,
\eta^{(3)}
+\frac{\partial {\cal S}^{(3)}}{\partial \xi^{(3)}}
(\xi^{(3)}, \eta^{(2)})
\right) \in {\bf R}^2 ,
}
\end{array}
\end{equation}
around 
$(\xi^{(3)}, \eta^{(3)}, \eta^{(2)}, \xi^{(4)})=0$,
we find the unique pair of functions, ${\tilde \xi}$ and
${\tilde \eta}$, in $\xi^{(4)}$ and $\eta^{(2)}$
which satisfy
$
\tau ({\tilde \xi}, {\tilde \eta}, \eta^{(2)}, \xi^{(4)})
=0
$
around $(\xi^{(4)}, \eta^{(2)})=0$. This shows (\ref{tilde-calS})
with (\ref{implicit-funct}).
We move on to the proof of (\ref{expansion-tilde-calS}) with
(\ref{o4}) in turn.
Equation (\ref{implicit-funct}) is expanded in $(\xi^{(4)},\eta^{(2)})$
as
\begin{equation}
\label{expansion-can-relation}
\eqalign{
{\tilde \xi}
=\xi^{(4)}
  +\frac{\partial {\cal S}_{4}}{\partial \eta^{(3)}}(\xi^{(4)}, \eta^{(2)})
   +o_{3}^{\prime}(\xi^{(4)}, \eta^{(2)}),
\\
{\tilde \eta}
=\eta^{(2)}
  +\frac{\partial {\cal S}_{3}}{\partial \xi^{(3)}}(\xi^{(4)}, \eta^{(2)})
   +o_{3}^{\prime \prime}(\xi^{(4)}, \eta^{(2)}),
}
\end{equation}
where $o_{3}^{\prime}$ and
$o_{3}^{\prime \prime}$ indicate the
terms of the order higher than $3$ in
$
\sqrt{\sum_{j=1}^{2}(\xi^{(4)}_j\phantom{}^2 +\eta^{(2)}_j \phantom{}^2)}
$.
Substituting (\ref{expansion-can-relation})
into (\ref{tilde-calS}),
we have (\ref{expansion-tilde-calS}). This completes the
proof.
\par\medskip\noindent
We are now in a position to see why the algorithm given here works well.
Lemma~A.1 and the setting-up (\ref{defeq-iterate}, \ref{cond-calS}),
are put together to imply that we have sloved through Steps-I,II
the inverse problem
\begin{equation}
\label{conv-H2-H4}
H^{(4)}(\xi^{(4)}, -\frac{\partial {\tilde {\cal S}}}{\partial \xi^{(4)}})
=
H^{(2)}(-\frac{\partial {\tilde {\cal S}}}{\partial \eta^{(2)}}, \eta^{(2)}),
\end{equation}
for $H^{(2)}$, where
${\tilde {\cal S}} \in \image D_{\xi^{(4)},\eta^{(2)}}$ up to degree-$4$.
This means that ${\tilde {\cal S}}(q,\eta)$ coincides up to degree-$4$
with the generating function $S(q,\eta)$ for the inverse problem for $G$,
so that the degree-$4$ inverse problem can be solved through Steps-I,II.
\end{document}